\documentclass[aps,prx,twocolumn,superscriptaddress,nofootinbib,
               preprintnumbers,longbibliography]{revtex4-2}

\usepackage{amsmath}
\usepackage{amssymb}
\usepackage{mathtools}
\usepackage{hyperref}
\usepackage{graphicx}
\usepackage{color}
\usepackage{comment}
\usepackage{url}
\usepackage{dcolumn}
\usepackage[caption=false]{subfig}
\usepackage{siunitx}
\usepackage[dvipsnames]{xcolor}

\definecolor{col1}{HTML}{1f77b4}
\definecolor{col2}{HTML}{ff7f0e}
\definecolor{col3}{HTML}{2ca02c}
\definecolor{col4}{HTML}{d62728}
\definecolor{col5}{HTML}{9467bd}
\definecolor{col6}{HTML}{8c564b}
\definecolor{col7}{HTML}{e377c2}
\definecolor{col8}{HTML}{7f7f7f}
\definecolor{col9}{HTML}{bcbd22}
\definecolor{col10}{HTML}{17becf}
\definecolor{col11}{HTML}{303030}

\graphicspath{
  {./}
}

\newcommand{\ben}{\begin{equation}}
\newcommand{\een}{\end{equation}}
\newcommand{\bea}{\begin{eqnarray}}
\newcommand{\eea}{\end{eqnarray}}
\newcommand{\ba}{\begin{array}}
\newcommand{\ea}{\end{array}}
\newcommand{\bit}{\begin{itemize}}
\newcommand{\eit}{\end{itemize}}

\newcommand{\half}{\frac12}

\newcommand{\bv}{\textbf{v}}

\newcommand{\cs}{c_\text{s}}

\begin{document}

\preprint{HIP-2021-29/TH}
  
\title{Decay of acoustic turbulence in two dimensions and implications \\ for
cosmological gravitational waves}

\newcommand{\HIPetc}{\affiliation{
Department of Physics and Helsinki Institute of Physics,
PL 64, 
FI-00014 University of Helsinki,
Finland
}}

\newcommand{\Sussex}{\affiliation{
Department of Physics and Astronomy,
University of Sussex, Falmer, Brighton BN1 9QH,
U.K.}}

\author{Jani Dahl}
\email[]{jani.dahl@helsinki.fi}
\HIPetc
\author{Mark Hindmarsh}
\email{mark.hindmarsh@helsinki.fi}
\HIPetc
\Sussex
\author{Kari Rummukainen}
\email{kari.rummukainen@helsinki.fi}
\HIPetc
\author{David J. Weir}
\email{david.weir@helsinki.fi}
\HIPetc

\date{August 17, 2022}

\begin{abstract}

\noindent
Gravitational waves from a phase transition associated with the generation of
the masses of elementary particles are within the reach of future space-based
detectors such as LISA. A key determinant of the resulting power spectrum,
not previously studied, is the lifetime of the acoustic turbulence which
follows. We study decaying acoustic turbulence using numerical simulations of a
relativistic fluid in two dimensions. Working in the limit of
non-relativistic bulk velocities, with an ultra-relativistic equation of state,
we find that the energy spectrum evolves towards a self-similar broken power
law, with a high-wavenumber behaviour of $k^{-2.08 \pm 0.08}$, cut off at very
high $k$ by the inverse width of the shock waves. Our model for the decay of
acoustic turbulence can be extended to three dimensions using the universality
of the high-$k$ power law and the evolution laws for the kinetic energy and the
integral length scale. It is used to build an estimate for the gravitational
wave power spectrum resulting from a collection of shock waves, as might be
found in the aftermath of a strong first order phase transition in the early
universe. The power spectrum has a peak wavenumber set by the initial length
scale of the acoustic waves, and a new secondary scale at a lower wavenumber
set by the integral scale after a Hubble time. Between these scales a
distinctive new power law appears. Our results allow more accurate predictions
of the gravitational wave power spectrum for a wide range of early universe
phase transition scenarios.
\end{abstract}

\maketitle

\section{Introduction}

\noindent The first direct observation of gravitational waves in
2015~\cite{LIGOScientific:2016aoc} started a new revolutionary era in
gravitational wave astronomy. For the first time, it was possible to make
observations without the limitations brought by detecting electromagnetic
radiation or particles. Gravitational waves travel at the speed of light, and
unlike electromagnetic radiation, interact extremely weakly with matter,
travelling mostly undisturbed through the universe, carrying with them
unfiltered information of their origins. These properties make them an
outstanding probe of the pre-recombination era
universe~\cite{Ricciardone:2016ddg}. Sources of gravitational waves in the very
early universe produce a stochastic gravitational wave
background~\cite{Christensen:2018iqi, Caprini:2018mtu} that could be detectable
with future gravitational wave detectors~\cite{Allen:1996vm, Maggiore:1999vm},
like the upcoming Laser Interferometer Space Antenna
(LISA)~\cite{LISA:2017pwj}.

One potential source of contributions to the stochastic gravitational
background of the very early universe is a first order cosmological phase
transition~\cite{Witten:1984rs,Hogan:1986qda,Kamionkowski:1993fg,
Caprini:2019egz}. Such transitions proceed via the nucleation, expansion, and
merger of bubbles containing the new low  temperature phase~\cite{Guth:1981uk,
Steinhardt:1981ct,Ignatius:1993qn, Espinosa:2010hh, Megevand:2017vtb}. The
phase transition comes to an end when all bubbles have merged with the
neighbouring bubbles so that the old phase has been replaced by the new one
everywhere in the fluid, leaving behind a characteristic spectrum of sound
waves~\cite{Hindmarsh:2013xza,Hindmarsh:2015qta,Hindmarsh:2017gnf,
Hindmarsh:2019phv} and if the transition is strong enough, significant
vorticity~\cite{Cutting:2019zws}. The sound waves are an important source of
gravitational waves. They persist in the fluid long after the phase transition
has completed, until dissipated away by the viscosity. Over time, these sound
waves can steepen into shock waves. Such a statistically random field of shocks
moving in various directions is known as acoustic turbulence~\cite{L_vov_1997,
Lvov:2000bdb}, and is the focus of this article.

Over the years the shock-containing compressional modes have received some
study but to a much lesser degree when compared to vortical turbulence. Perhaps
the most famous of such studies is that of the relatively simple Burgers'
equation~\cite{Frisch:2000td, BEC_2007}, which shares many of the properties
seen in the Navier-Stokes equations, apart from the chaotic behaviour and
randomness rising from small perturbations in the initial conditions. This is
because it is possible to integrate Burgers' equation explicitly. Burgers'
equation appears in the asymptotic limit in many physical situations, and has
been extensively studied due to its simplicity.

As for the Navier-Stokes equations, there have been some earlier studies that
deal with the compressional modes in fluids with a non-relativistic equation of
state. Numerical simulations of the two- and three-dimensional Navier-Stokes
equations with a longitudinal velocity component were performed by Porter,
Pouquet and Woodward in Ref.~\cite{Porter1992ANS} in the supersonic limit. They
pay attention to the power laws seen in the energy spectra and the kinetic
energy fractions between the longitudinal and transverse modes. The
interactions between the compressible and rotational modes in a
three-dimensional case were studied in 1990's in
Refs.~\cite{Kida1992EnergyAS,DUCROS1999517} with resolutions up to $1024^3$. Of
the more cosmologically oriented papers using relativistic fluid equations, one
worth highlighting is a paper by Pen and Turok~\cite{Pen:2015qta} that contains
one-, two- and three-dimensional simulations of shock formation in primordial
acoustic oscillations. 

In this paper we study two-dimensional decaying acoustic turbulence using
numerical simulations with relativistic fluid equations and random initial
conditions. The emphasis is  on the profile of the generated shock waves, their
effect on the shape of the energy spectrum, and the decay properties of the
kinetic energy and the integral length scale. Using the obtained results, we
also make an estimate for the gravitational wave power spectrum resulting from
shocks in a three-dimensional fluid flow.

We have chosen to conduct the simulations in two dimensions for several
reasons. The most important of these is that based on the existing literature,
the shocks -- amongst other phenomena -- have the same properties, like the
inertial range power laws in the energy spectrum, in two and three dimensions.
However, the two-dimensional case is simpler to analyse: for example, it is
easier to locate the shocks in two dimensions; some quantities, like the
vorticity, are simpler (it being a scalar in 2D); and there are additional
conserved quantities compared to 3D. In addition, there is the clear advantage
of 2D being more computationally efficient, allowing for the use of larger grid
sizes, increasing the dynamic range of the simulations. This makes it easier to
study non-linear phenomena like turbulence and shocks. In 3D, the largest
simulations to date have  lattice sizes of $4200^3$~\cite{Hindmarsh:2017gnf},
and have not yet simulated sufficiently fast fluid flows for long enough to
show the development of turbulence after a cosmological phase transition. 
Here, we simulate on grid sizes up to $10000^2$, for long enough to  easily see
the development and decay of shocks.
  
We also save some compute time by starting simulations with the velocity and
density perturbations as a random field with given power spectra, rather than
simulating the whole phase transition. This allows us to conclude that the
effects we observe are not special to phase transitions. This is similar to the
approach of Ref.~\cite{RoperPol:2019wvy},  studying gravitational wave
production by vortical turbulence, which starts its simulations with the
Kolmogorov spectrum.

The contents of this article are as follows: Section~\ref{Methods} contains
information about the fluid equations, details of the numerical simulations and
the initial conditions, and lists some useful quantities used in characterising
the state of the fluid. Section~\ref{Results} concerns the results of our
numerical simulations and is divided into several subsections. In
Section~\ref{shockshape}, an analytical form for the shock shape is derived
using the fluid equations. Section~\ref{EnSpec} focuses on the energy spectrum
and its evolution over time, and in~\ref{kinendec} the decay of kinetic energy
and the integral length scale is studied. The last
subsection,~\ref{transenerg}, takes a closer look at the transverse kinetic
energy that arises from the longitudinal only initial conditions under these
fluid equations. In Section~\ref{GravSpec} an estimate is built for the
gravitational wave power spectrum resulting from a collection of sound waves
seen in our simulations. Two appendices are also included, Appendix~\ref{AppA}
being about testing the results obtained in Section~\ref{shockshape} by
conducting runs in a shock tube. Appendix~\ref{AppB} provides a more in depth
look at the initial conditions and some more technical aspects of the
simulations. Also listed are the runs used to obtain the tables and figures
presented in this paper, and the initial conditions for each of the runs. In
this paper we take the speed of light $c=1$.

\section{Methods} \label{Methods}

\noindent In this paper we study the evolution and properties of
two-dimensional decaying acoustic turbulence using numerical simulations of a
relativistic fluid. The equations we have employed are obtained from the
relativistic fluid equations by expanding them to second order in first
order small quantities that we have taken to be the non-relativistic bulk
velocity $\bv$, and the bulk and shear viscosities. We also relate the
pressure $p$ and energy density $\rho$ via the ultra-relativistic
equation of state $p = c_s^2 \rho$, where $c_s$ is the speed of sound
parameter, which has the value $1/\sqrt{3}$ in the case of a radiation fluid.
The derivation of the inviscid part of these equations is discussed in
more detail in~\cite{Brandenburg:1996fc}. They can be written as
\begin{widetext}
\begin{align}
\frac{\partial \rho}{\partial t} + (1+c_s^2) \nabla \cdot (\rho \bv) &
= 0 \label{cont} \\
\frac{\partial \bv}{\partial t} + \bv \cdot \nabla \bv - c_s^2 \bv
(\nabla \cdot \bv) + \frac{c_s^2}{\rho(1+c_s^2)} \nabla \rho & = 
\frac{1}{1+c_s^2} \left[ \eta \nabla^2\bv + 
\left(\frac{1}{3} \eta + \nu \right) \nabla (\nabla \cdot \bv) \right] \, , 
\label{EulerEq}
\end{align}
\end{widetext}
where $\eta$ and $\nu$ are the kinematic shear and bulk viscosity respectively.
They enter the equations via the additions of the anisotropic stress tensor and
the viscous bulk pressure to the energy momentum tensor. In the very early
universe the Reynolds number is expected to be large and the shear viscosity to
be dominant; its magnitude can be expressed in terms of the temperature and the
electromagnetic gauge coupling parameter~\cite{Arnold:2006fz}. With our choice
of scheme, viscosity is required to keep the numerical solution stable in cases
where there is significant power in the longitudinal modes. In the longitudinal
case shear and bulk viscosities also act in effectively the same way. The lack
of an external forcing term means that the kinetic energy in the system decays
over time, as it is being dissipated into internal energy by the viscosity at
small length scales.

We define the spectral density $P(k)$ through the two-point correlation
function of a homogenous and isotropic velocity field as
\begin{equation}
\left\langle v_i (\mathbf{k}) v_i (\mathbf{k}^\prime) \right\rangle = (2 \pi)^2
P(k) \delta(\mathbf{k} - \mathbf{k}^\prime)
\end{equation}
with $v_i(\mathbf{k})$ being the Fourier components of the velocity, related
through the Fourier transform pair
\begin{align}
v_i(\mathbf{k}) &= \int  v_i(\mathbf{r}) e^{-i \mathbf{r} \cdot \mathbf{k}} \,
d^2r \\
v_i(\mathbf{r}) &= \frac{1}{(2 \pi)^2}\int v_i(\mathbf{k}) e^{i \mathbf{r}
\cdot \mathbf{k}} \, d^2k \, ,
\end{align}
where $\mathbf{k}$ is the wave vector. We define a quantity $E(k)$ through 
\begin{equation} \label{PowSpecEnSpecRel}
E(k) = \frac{k}{4 \pi} P(k) \, ,
\end{equation}
where $P(k)$ is the spectral density, such that 
\begin{equation} \label{En_spec_def}
\half \left\langle \bv^2 \right\rangle = \int\limits_0^{\infty} E(k) \, dk \, ,
\end{equation}
from which we directly obtain the root mean square (rms) value of the
velocity vector field.
In a system with a non-relativistic equation of
state, $E(k)$ is also the linear spectrum of the kinetic energy per unit
mass, so we will refer to it as the energy spectrum.  The true specific kinetic
energy in our system is 
$(1 + \cs^2) \left\langle \bv^2 \right\rangle$.

The velocity field is decomposed into longitudinal and transverse components so
that
\begin{equation}
\bv = \bv_\parallel + \bv_\perp \, ,
\end{equation}
where the components fulfil the properties
\begin{equation}
\nabla \cdot \bv_\perp = 0 \, , \qquad \nabla \times \bv_\parallel = 0 \, .
\end{equation}
This decomposition also splits the energy spectrum into two parts,
$E(k) = E_\parallel(k) + E_\perp (k)$, where the longitudinal spectrum contains
the contribution from acoustic turbulence, and the transverse part the vortical
contributions associated with traditional fluid turbulence that consists of
vortices of various sizes.

The fluid equations are integrated numerically using finite difference methods.
Time integration is performed using the fourth order Runge-Kutta scheme, and
spatial derivatives are evaluated with a second order central difference
scheme.\footnote{We have checked that no significant changes to our results
would be introduced by a fourth order scheme. For more information about
the scheme and its viability, see Appendix~\ref{AppA}.} The
spatial grid is a square with $N^2$ points and unit spacing. The time step size
is chosen as $\Delta t = 0.2 \Delta x$, providing a stable numerical solution.
Periodic boundary conditions are enforced on all edges of the grid. The
corresponding reciprocal lattice is spanned by the wave vectors $\mathbf{k}_i$,
whose elements obtain values in the range $[- \pi, \pi)$ with a spacing of
$\Delta k = 2 \pi /(N \Delta x)$.

The initial conditions are given for the longitudinal and transverse components
using an initial spectral density of the form
\begin{equation} \label{initPowSpec}
P(k) = A \frac{
  (k/k_p)^{\beta_0}}{\left[ 1 + (k/k_p)^{\alpha_0/ \gamma} \right]^\gamma}
e^{-(k/k_d)^2} \, .
\end{equation}
The parameters $\alpha_0$ and $\beta_0$ set the initial values for the inertial
range and low-$k$ power laws, $\gamma$ affects the shape around the peak of the
spectrum, and $k_p$ is the initial wavenumber around which the peak is located.
The inverse of $k_p$ deterimines the integral length scale $L$ characterizing
the length scale at which most of the energy is located. We have also
introduced an exponential suppression factor in order to reduce discretization
effects at high wavenumbers. This suppression is controlled by the parameter
$k_d$. Since we are interested in acoustic turbulence only, we set
$P_\perp(k)=0$ initially, i.e.~the initial velocity field is purely
compressible. The Fourier components obtained from $P_\parallel(k)$ are given
random phases in such a way that the resulting initial velocity field is real
and statistically random. The energy density is initialized by writing it out
as
\begin{equation}
\rho (\mathbf{r}, t) = \rho_0 + \delta \rho (\mathbf{r}, t) \,
\end{equation}
where the density perturbation $\delta \rho$ is initialized in the same way as
the velocity components. More information about the execution of the
simulations and their initial conditions can be found in Appendix~\ref{AppB}.

Next, we define some quantities that are useful in analysing the flows. We
write the rms velocity of the longitudinal component as 
$\bar{v}=\sqrt{\left\langle \bv_\parallel^2 \right\rangle}$, and define the
integral length scale of the longitudinal component as 
\begin{equation} \label{int_len_scale}
L = \frac{2}{\bar{v}^2} \int\limits_0^\infty \frac{1}{k} E_\parallel(k) \, dk
\, .
\end{equation}
From the initial values of these two quantities we can define a time scale,
\begin{equation}
t_s=L_0/\bar{v}_0
\end{equation}
where $\bar{v}_0$ is the initial value of the rms longitudinal velocity, and
$L_0$ is the initial value of the integral scale. In addition to the integral
scale, there are other relevant length scales constructed from the effective
viscosity $\mu = 4 \eta/3 + \nu$, the rms velocity $\bar{v}$, and the quantity
\begin{equation}
  \label{eq:dilatationsquared}
  \mathcal{D} = \left\langle (\nabla \cdot \bv)^2 \right\rangle,
\end{equation}
the compressional part of the enstrophy, which can be used to quantify
``shockiness'' in the system. First, we have 
\begin{equation} \label{swidth}
\delta_s = \mu/\bar{v} \, , 
\end{equation}
which we shall see characterises the shock width. We also have the longitudinal
counterparts of the Kolmogorov and Taylor microscales $L_K$ and $L_T$.  We
define the Kolmogorov microscale as
\begin{equation}
L_K = \left( \frac{\mu^3}{\epsilon} \right)^{1/6} \, ,
\end{equation}
where $ \epsilon = - \dot{\mathcal{D}}$ is the dissipation rate of
$\mathcal{D}$. From the equations of motion it follows that 
\begin{equation}
\epsilon 
= \frac{4 \mu}{1+c_s^2} \int\limits_0^\infty k^4 E_\parallel (k) \, dk \, .
\end{equation}
As in the case of vortical fluid turbulence, the Kolmogorov microscale
specifies the length scale at which viscosity is dominant and dissipates
kinetic energy into internal energy. The Taylor microscale is an intermediate
length scale located between the integral and Kolmogorov length scales at which
viscous effects become significant, and is defined by
\begin{equation}
L_\text{T} = \sqrt{ \frac{ \bar{v}^2 } {\mathcal{D}} } \, .
\end{equation}
The Taylor and Kolmogorov wavenumbers are defined as inverses of the
corresponding length scales. We also define the longitudinal counterpart of the
Reynolds number
\begin{equation} \label{ReynoldsNum}
\text{Re} = \frac{\bar{v} L}{\mu} \, ,
\end{equation}
that, as in the vortical only case, characterises the strength of non-linear
effects in the flow, which in the longitudinal case means shocks. In other
words, large values of the longitudinal Reynolds number lead to a formation of
very strong and sharp shocks.

\section{Results} \label{Results}

\noindent We have performed numerical simulations of acoustic turbulence with
grid sizes of $N=4080$ and $N=10080$ with various initial power spectra
(\ref{initPowSpec}) for the longitudinal component, leading to various shock
formation times, and initial longitudinal Reynolds numbers in the range 16-223.
We call runs with an initial Reynolds number that lies at the end of this range
high Reynolds number runs. These kind of runs are obtained by increasing the
initial rms velocity and also by increasing the initial integral length scale,
which moves the top of the energy spectrum to lower wavenumbers. We have run
for about 60 shock formation times in all of our runs to give the system enough
time to show sufficient decay characteristics. A table listing each run and
their initial conditions is found in Appendix~\ref{AppB}. In this section we
shall present our findings from these runs focusing on the shape of the shocks,
their impact on the energy spectrum, the decay of the longitudinal kinetic
energy and the integral length scale, and the generation of transverse kinetic
energy under these equations from the longitudinal only initial conditions.

\subsection{Shock shape} \label{shockshape}

\noindent In order to study the shape of the shock waves, we solve equations
(\ref{cont}) and (\ref{EulerEq}) for a single shock moving towards the positive
x-axis using the ansatz
\begin{equation}
\rho (\mathbf{r}, t) = L(k_s(x-ut)), \quad \vec{v} (\mathbf{r}, t) = 
V(k_s(x-ut)) \vec{e}_x \, .
\end{equation}
Here $u$ denotes the shock velocity. The resulting differential equation is
then written in terms of $\chi=k_s(x-ut)$ for $V(\chi)$ and is simplified by
assuming that $V(\chi) \ll u$. Its solution is
\begin{equation} \label{ShockProf}
V(x, t) = \frac{\sqrt{a^2+2bC}}{b} \tanh \left[ k_s (x-x_0-ut) \right]
- \frac{a}{b} \, ,
\end{equation}
where the parameters $a$ and $b$ can be written as:
\begin{gather}
a = u \left(1 - \frac{c_s^2}{u^2} \right) \\
b = (1+c_s^2) \left( \frac{c_s^2-1}{c_s^2+1} - \frac{c_s^2}{u^2} \right) \, .
\end{gather}
The integration constant $C$ is fixed using the conditions that on the right
side of the shock $V$ approaches the value $V_+$, and on the left side the
value $V_-$ while the derivative of $V$ approaches zero on both sides. For a
right-moving shock we also have $V_- > V_+$. These conditions fix $C$ as
\begin{equation}
C = a V_+ + \frac{b}{2} V_+^2 = a V_- + \frac{b}{2} V_-^2 \, .
\end{equation}
The shock velocity is solved from this equation and can be written in the form
\begin{equation} \label{ShockVel}
u = c_s \left( 1 - \frac{2}{1+c_s^2} \xi \right)^{- \frac{1}{2}} \, , \quad
\xi = \left( 1 + \frac{2}{\widetilde{\delta \rho}_+
+ \widetilde{\delta \rho}_-} \right)^{-1}
\end{equation}
where $\widetilde{\delta \rho}_+$ and $\widetilde{\delta \rho}_-$ are the
values of the fractional density perturbation on the left and right sides of
the shock, obtained by replacing $V_+$ and $V_-$ using the relation between $V$
and the fractional density perturbation
\begin{equation} \label{VelDenRel}
V(x,t) = \frac{u}{1+\cs^2} \widetilde{\delta \rho} (x, t) \, , \quad
\widetilde{\delta \rho} (x, t) = \frac{\delta \rho(x,t)}{\rho_0}.
\end{equation}
Equation (\ref{ShockVel}) gives us the expected result of the shock velocity
always being larger than the speed of sound for a right-moving shock with
$\widetilde{\delta \rho}_+=0$, since the smallest value $\xi$ can obtain is
zero in the limit of the shock wave amplitude going to zero. It is also evident
that the shock velocity increases with increasing amplitude.

The width of the shock is controlled by the parameter $k_s$ that can be written
in terms of the above quantities as
\begin{equation}
k_s = \frac{3 (1+c_s^2) \sqrt{a^2+2bC}}{8 \mu}
\end{equation}
and whose inverse is of the same order of magnitude as the shock width
$\delta_s$.
The parameters $a$, $b$ and $C$ all increase with increasing shock velocity,
which indicates that steep shocks are obtained when the amplitude of the shocks
is large. Here the effects of the viscosities are clearly seen, with small
viscosity values leading to steep shocks. We have conducted shock tube runs to
study and verify the results obtained here by investigating shocks in a very
narrow and long grid. These are discussed in Appendix~\ref{AppA}.

\begin{figure*}
\centering
\subfloat[]{
\includegraphics[width=0.5\textwidth]{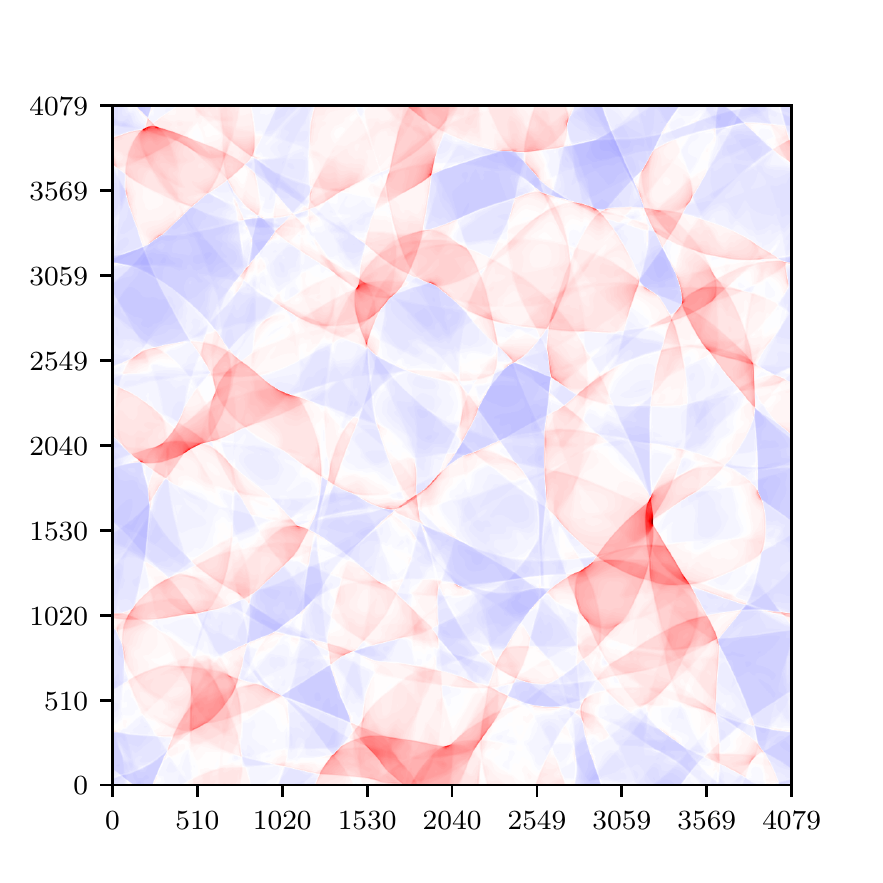}
\label{fig:fig1a}
}
\subfloat[]{
\includegraphics[width=0.5\textwidth]{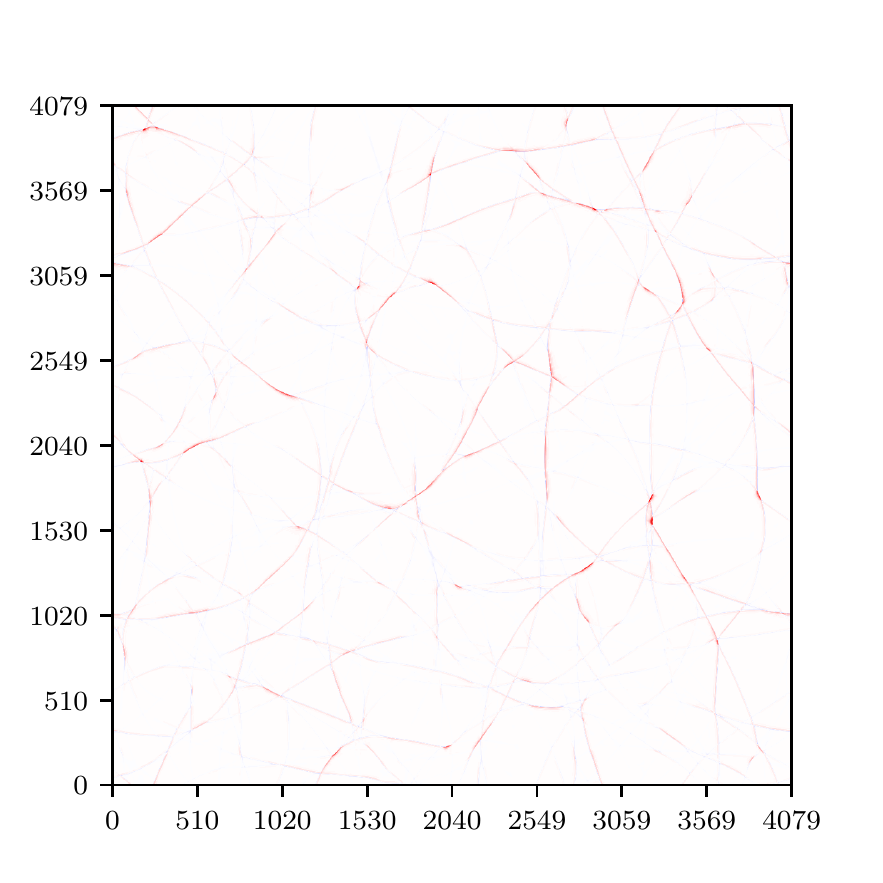}
\label{fig:fig1b}
}
\caption{The density perturbation $\delta \rho$ of a $4080^2$
  resolution run that has the same initial conditions as run 9 (a),
  and the divergence of the corresponding velocity field $\nabla \cdot
  \bv$ (b), showing the locations of the shocks at $t \approx 6.5 t_s$.}
\label{fig:rhodiv}
\end{figure*}

\begin{figure*}
\centering
\subfloat[]{
\includegraphics[width=0.5\textwidth]{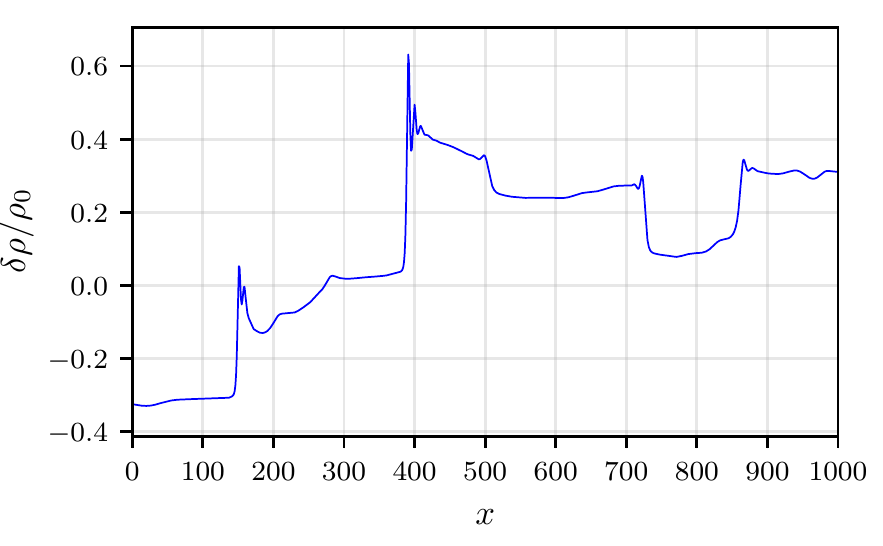}
\label{fig:fig2a}
}
\subfloat[]{
\includegraphics[width=0.5\textwidth]{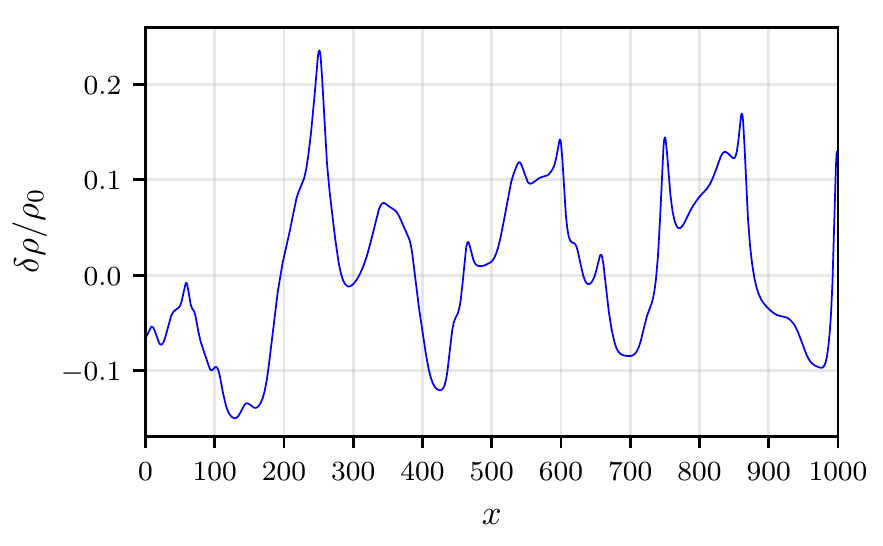}
\label{fig:fig2b}
}
\caption{Zooms of the fractional density perturbation slices of the same
$4080^2$ resolution run as in Figure~\ref{fig:rhodiv} (a), and a similar
moderate Reynolds number run (b) that has identical initial conditions to those
of run 2. Both figures show the shocks after about $6.5$ shock formation
times.}
\label{fig:shockslices}
\end{figure*}

In our 2D simulations the initially smooth density and velocity fields
generate multiple shock waves moving in various directions, after a time of
order $t_s$. This can be seen in Figure~\ref{fig:rhodiv}, which on the left
shows a contour plot of the density perturbation shortly after the shocks have
formed. In the second plot on the right, the divergence of the velocity field
has been plotted to highlight the shocks. Figure~\ref{fig:shockslices} shows
zoomed in slices of the fractional density perturbation both in the high and
low Reynolds number cases. In the case of the former, oscillations can be seen
near the top of the shock, similar to the Gibbs phenomenon~\cite{Gibbs}. This
limits the obtainable Reynolds numbers, as reducing the viscosity too much
causes these oscillations to grow, eventually ruining the solution. This also
has an effect on the shape of the energy spectrum around the Kolmogorov
microscale, creating a bump in the spectrum at this wavenumber range.

\subsection{Shocks and the energy spectrum} \label{EnSpec}

\noindent Based on our simulations that use various different initial spectral
densities (\ref{initPowSpec}), we find that as the initial conditions steepen
into shocks, the features induced by the initial conditions near the peak of
the energy spectrum are erased, and that the energy spectrum obtains a
universal broken power law form whose power law values differ from those of
the initial conditions. After a single shock formation time, the inertial range
power law located between the integral length scale and the Taylor microscale
settles into the well-known value of -2, first proposed and obtained by
Burgers~\cite{Burgers_1948} for the one-dimensional Burgers equation, and later
generalised to multiple dimensions in the case of the Euler and the continuity
equation by Kadomtsev and Petviashvili~\cite{KP}.

\begin{figure}
\begin{center}
\includegraphics[width=\columnwidth]{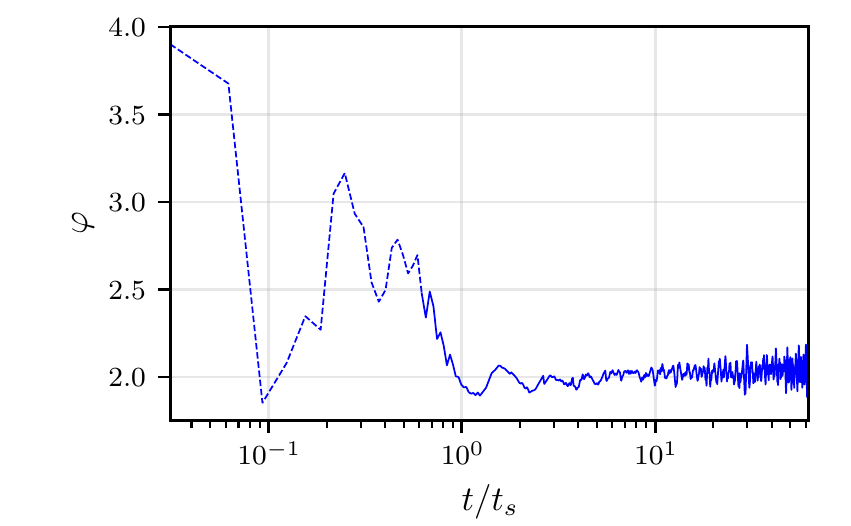}
\end{center}
\caption{\label{fig:inerpowlaw} Evolution of the inertial range power law
index obtained by fitting a power law $k^{- \varphi}$ to the data of run 9. The
bounds of the fit are the Taylor and half times the Kolmogorov wavenumbers when
$t/t_s<0.6$ (dashed curve) and the integral wavenumber and the Taylor
wavenumber otherwise.}
\end{figure}

The evolution of the inertial range power law in one of our runs has been
plotted in Figure~\ref{fig:inerpowlaw} as a function of the number of shock
formation times. Due to strong oscillations in the spectrum at early times,
the data for the plot has been obtained by fitting a power law $k^{- \varphi}$
to two intervals; between the Taylor and half the Kolmogorov wavenumber at
early times when $t/t_s < 0.6$, and between the integral wavenumber and the
Taylor wavenumber otherwise. Early on, obtaining decent fits of the inertial
range is obstructed by these oscillations, so we have tracked the evolution of
the power law range of the initial conditions instead, which initially develops
towards a similar power law value but at a higher wavenumber range. At
$t/t_s = 0.6$ the oscillations have weakened and the two ranges coincide, to a
reasonable accuracy, so we have opted to change the limits of the fit at this
particular time. 

\begin{figure*}
\centering
\begin{minipage}[t]{1.0\columnwidth}
\centering
\includegraphics[width=\columnwidth]{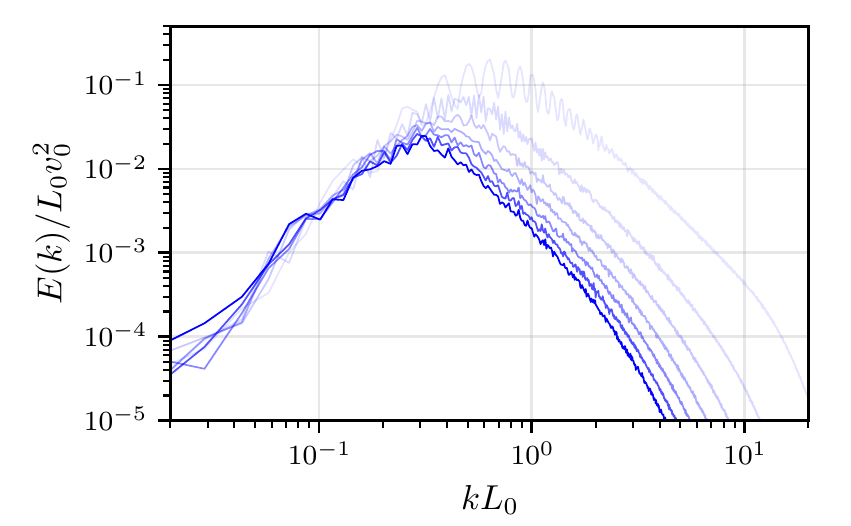}
\caption{\label{fig:energyspecs} The scaled energy spectrum of run 2 plotted
every $10 t_s$ from $3 t_s$ onwards. Dark colours correspond to late times.}
\end{minipage} \hfill
\begin{minipage}[t]{1.0\columnwidth}
\centering
\includegraphics[width=\columnwidth]{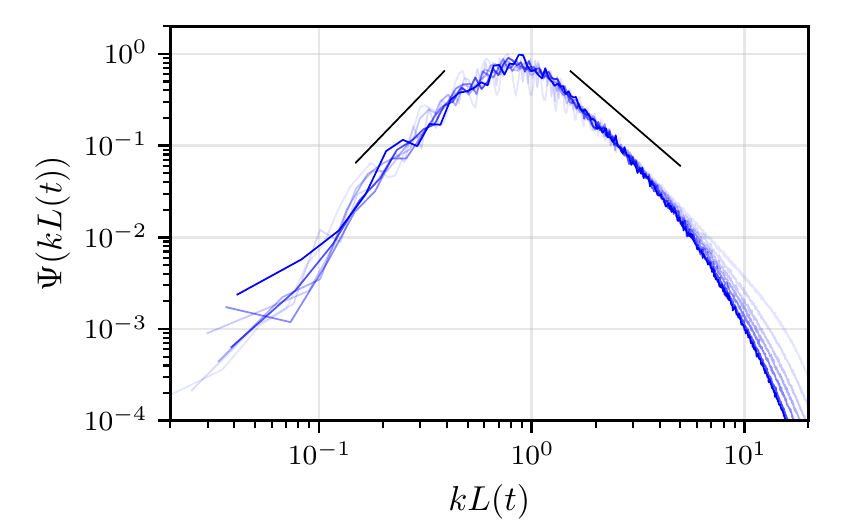}
\caption{\label{fig:collapseplots} The energy spectra of
Figure~\ref{fig:energyspecs} collapsing into the function $\Psi(Lk)$. The black
line above the inertial range describes the $k^{-2}$ power law, while the black
line above the low-$k$ range goes like $k^{2.4}$.}
\end{minipage}
\end{figure*}

In order to study and determine the universal shape of the spectrum, we extract
the time dependence from the spectrum. Figure~\ref{fig:energyspecs} shows the
time evolution of the spectrum with dark lines corresponding to late times.
Over time the integral length scale $L$, introduced in section~\ref{Methods},
increases as evidenced by the shift of the peak of the spectrum towards small
wavenumbers. Thus we fix the location of the peak by scaling the wavenumber by
$L$, so that the spectrum becomes a function of $\kappa = L(t)k$. The other
time dependent feature of the spectrum is the decay of the compressional
kinetic energy $\mathcal{E}_\parallel$, which causes the magnitude of the
spectrum to decrease. Thus, we write the spectrum in the form
\begin{equation} \label{EnPsi}
E_{\parallel} (\kappa, t) = L(t) \mathcal{E}_{\parallel}(t) \Psi(\kappa) \, ,
\quad \kappa = L(t)k \, .
\end{equation}
The function $\Psi(\kappa)$ is plotted in Figure~\ref{fig:collapseplots} at
even time intervals after the shocks have formed, and we see the spectra
collapsing onto a single function on all but the very smallest length scales.

\begin{figure}
\begin{center}
\includegraphics[width=\columnwidth]{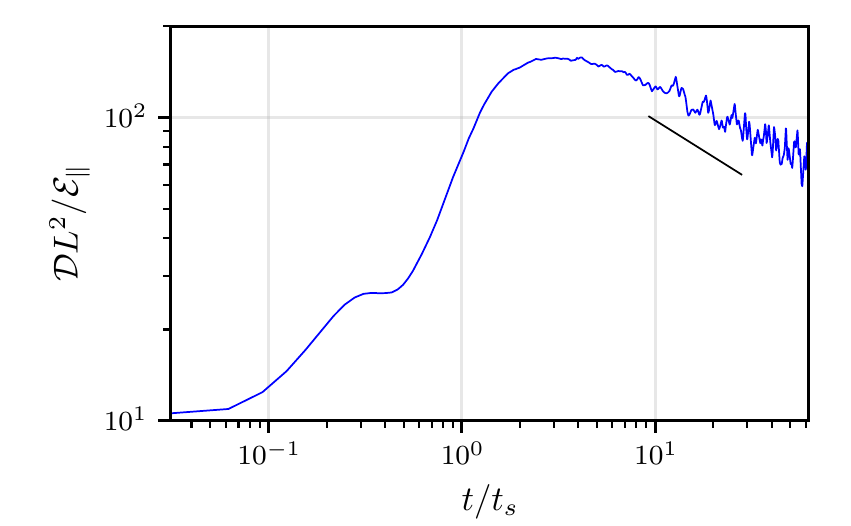}
\end{center}
\caption{\label{fig:avg_div_sqrd} Plot of the dimensionless quantity
$\mathcal{D} L^2 / \mathcal{E}_\parallel$ that is used to measure shockiness of
the system as a function of shock formation times. The data for the plot has
been obtained from run 9. The black line indicates a $t^{-0.4}$ power law.}
\end{figure}

In the range where the spectra collapse well, we model the function $\Psi$ by
a broken power law form. We assume that this form holds at wavenumbers
corresponding to length scales larger than the Taylor microscale, so that
\begin{equation} \label{BrokPowLaw}
\Psi(\kappa) = \Psi_0 \frac{(\kappa/ \kappa_p)^{\beta}}{
  1+(\kappa/ \kappa_p)^\alpha} \, , \quad \kappa \ll L/L_T \, ,
\end{equation}
where $\beta$ is the low-$k$ power law index, and the inertial range
power law is given by $\beta-\alpha$. From equation (\ref{EnPsi}) it follows
that the integral of $\Psi$ over all values of $\kappa$ must equal unity, and
another condition follows from substituting (\ref{EnPsi}) into the definition
of the integral length scale in Equation (\ref{int_len_scale}). For both of
these conditions to be satisfied simultaneously, the parameters $\Psi_0$ and
$\kappa_p$ must fulfil
\begin{equation} \label{Afix}
\Psi_0 = \frac{\alpha}{\pi}  \sin \left(\frac{\pi \beta}{\alpha}\right)
\end{equation}
and
\begin{equation} \label{KappaFix}
\kappa_p =  \frac{ \sin \left(\frac{\pi  (\beta+1)}{\alpha}\right)}{ \sin
\left(\frac{\pi  \beta}{\alpha}\right)} \, ,
\end{equation}
when $\beta - \alpha < -1$, meaning that these parameters get fixed by the
normalisation condition and the choice of $L$ as the integral scale. In the
high-$\kappa$ region where the collapse is not as good, and the function still
changes a little over time. We ascribe this temporal behaviour to the changing
shape of the shocks caused by the viscous dissipation. Thus, we expect the
function $\Psi$ to be a broken power law modulated by a function that depends
on the width of the shocks. To quantify the dilatation of the shocks, we use
the dimensionless quantity $\mathcal{D} L^2 / \mathcal{E}_\parallel$ to measure
shockiness in the system, where $\mathcal{D}$ is defined through
Eq.~(\ref{eq:dilatationsquared}). The quantity $\nabla \cdot \mathbf{v}$, often
called the dilatation, obtains large values at the locations of the shock waves
briefly after shock formation in comparison to the values seen in the initial
conditions, which leads to an increase in its rms value $\sqrt{\mathcal{D}}$.
The dimensionless quantity is plotted in Figure~\ref{fig:avg_div_sqrd}, and a
sharp increase in its value can be seen around one shock formation time, after
which the quantity decreases, as the shocks deteriorate.

\begin{figure}
\begin{center}
\includegraphics[width=\columnwidth]{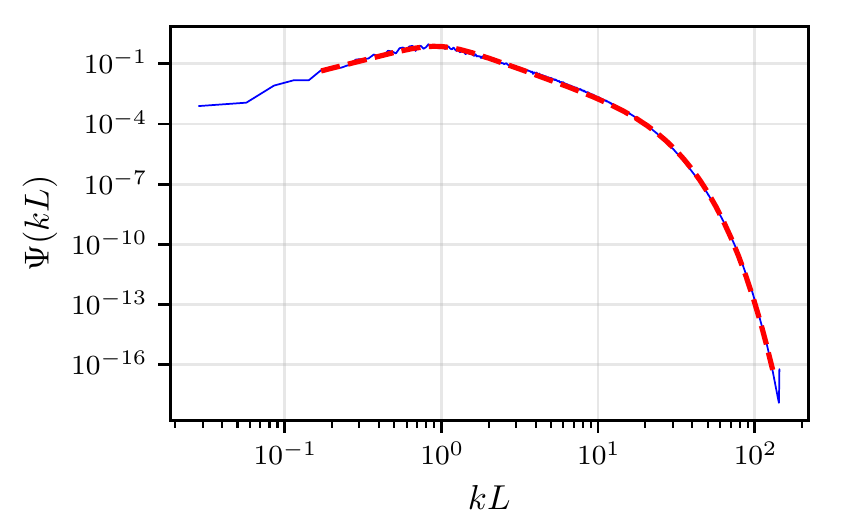}
\end{center}
\caption{\label{fig:Psifit} The function $\Psi(\kappa)$, where the blue line
is the curve obtained from simulation data of run 2 at $t \approx 20 t_s$, and
the dashed red line is a fit using Equation~(\ref{CollapsePsiEq}). The obtained
values for the fit parameters are $\widetilde{\Psi}_0=0.0034$, $\alpha \approx
4.801$, $\beta \approx 2.013$. $\kappa_p \approx 0.976$, and
$\kappa_s=12.541$.}
\end{figure}

In order to determine what impact the shocks have on the energy spectrum, we
follow the method presented in Ref.~\cite{Kuznetsov} to find the form of the
two-dimensional energy spectrum using the one-dimensional spectrum. The
one-dimensional energy spectrum of a tanh shock is obtained using the Fourier
transform and has the form
\begin{equation} \label{1dspec}
E_1(k) = |\mathcal{F}(\tanh(k_s x))|^2 = \frac{\pi^2}{ k_s^2} \text{csch}^2
\left( \frac{\pi k}{2 k_s} \right) \, ,
\end{equation} 
which can be related to the $D$-dimensional spectrum by separating the
wavevector $\mathbf{k}$ into two parts; $\mathbf{k}_1$ and its transverse
projection $\mathbf{k}_\perp$ and then by integrating over the latter
\begin{align}
E_1(k_1) &= \int E_D (|\mathbf{k}|) d \mathbf{k}_\perp \\
&= \frac12 \Omega_{D-1} \int\limits_{k_1^2}^\infty E_D(s) (s^2 - k_1^2)^{
  \frac{D-3}{2}} \, ds^2 \, .
\end{align}
where $\mathbf{k} = \mathbf{k}_\perp + \mathbf{k}_1$ and $\Omega_{D-1}$ is the
solid angle of the ($D-2$)-sphere. For $D=2$ the equation can be written as
\begin{equation}
E_1(k_1) = \int\limits_{k_1^2}^\infty \frac{E_2(\sqrt{u})}{\sqrt{u-k_1^2}} \,
du \, .
\end{equation}
Now we can use the property, that $E_1$ is a first order Liouville-Weyl
fractional integral~\cite{Erdelyi} of $E_2$ to solve for the two-dimensional
spectrum, yielding
\begin{equation} \label{E2LW}
E_2(k) = \frac{1}{\pi} \int\limits_{k^2}^\infty \frac{1}{\sqrt{u-k^2}}
\frac{d}{du} E_1(\sqrt{u}) \, du \, .
\end{equation}
Substituting Equation (\ref{1dspec}), changing the variable, and defining
\begin{equation}
E_2 = \frac{\pi^2}{k_s^3} \mathcal{I} \, , \quad P= \frac{\pi k}{2 k_s}
\end{equation}
allows us to write (\ref{E2LW}) as
\begin{equation} \label{IP}
\mathcal{I}(P) = \int\limits_1^\infty \frac{ds}{\sqrt{s^2-1}}
\frac{\cosh(P s)}{\sinh^3(P s)} \, .
\end{equation}
The integral in $\mathcal{I}$ does not have a closed form solution, but its
asymptotic behaviour at small and large values of the argument can be found to
be
\begin{equation}
\mathcal{I} (P) \sim
\begin{cases}
\dfrac{\pi}{4 P^3} \, , \quad P \ll 1 \\
\\
2 \sqrt{\dfrac{\pi}{P}} e^{-2P} \, , \quad P \gg 1
\end{cases} \, .
\end{equation}
We now propose the function $\Psi(\kappa)$ to have the form
\begin{equation}
\Psi(\kappa) = \widetilde{\Psi}_0 \frac{(\kappa/ \kappa_p)^{\beta+3}}{
  1+(\kappa/ \kappa_p)^\alpha} \mathcal{I}
  \left( \frac{\pi \kappa}{2 \kappa_s} \right) \, , \label{CollapsePsiEq}
\end{equation}
where $\kappa_s=k_s L$. Note that the parameter $\beta$ still denotes the
low-$k$ power law. Figure~\ref{fig:Psifit} shows $\Psi(\kappa)$ obtained from
simulation data in comparison to the fit resulting from using the equation
above. It is seen that the fit is very good in the high-$\kappa$ region. At the
wavenumber range between the Taylor and the Kolmogorov wavenumbers, the fit
deviates a bit from the simulation data, leading to slightly steeper values for
the inertial range power than $k^{-2}$. The fit in this range can be improved
by increasing the complexity of the fitting function, for example, by using a
double broken power law instead, but for our purposes we deem
Equation~(\ref{CollapsePsiEq}) to be a good enough estimate for the spectral
collapse function $\Psi$.

\subsection{Decay of longitudinal kinetic energy} \label{kinendec}
\begin{figure}
\begin{center}
\includegraphics[width=\columnwidth]{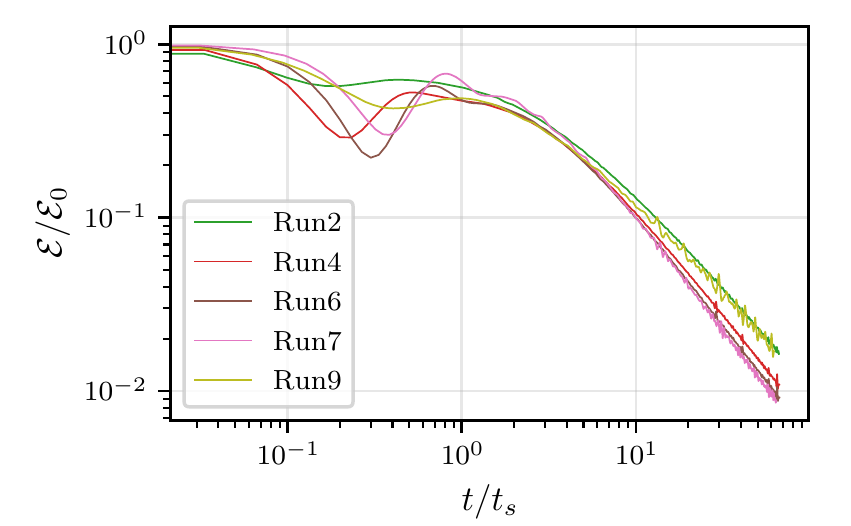}
\end{center}
\caption{\label{fig:kin_en} The kinetic energy normalised by the
  initial value plotted for some the runs listed in
  Table~\ref{tab:table3} as a function of the shock formation time.}
\end{figure}

\noindent Energy is dissipated into heat by the viscosity at small length
scales, and since our fluid equations do not contain a forcing term, the total
kinetic energy decreases over time. Figure~\ref{fig:kin_en} plots the kinetic
energy normalised by its initial value for several runs as a function of the
number of shock formation times. It is seen that after about 10 shock formation
times the kinetic energy decays following a power law form.
In order to find an analytical function that models the kinetic energy
behaviour seen in the figure, we have applied the analysis made by Saffman in
Ref.~\cite{saffman1971}, but to the longitudinal-only case instead. The
starting point is the relation describing the kinetic energy decay due to
viscous dissipation, which for the fluid equations (\ref{cont}) and
(\ref{EulerEq}) can be shown to be 
\begin{equation} \label{viscousdiss}
\half \frac{d \left\langle \bv^2_\parallel \right\rangle}{dt} = - \frac{2 \mu}{1+\cs^2} \int\limits_0^\infty k^2 E_\parallel(k) \, dk \, ,
\end{equation}
when the vorticity $\nabla \times \mathbf{v}$ is zero. There is some
vorticity generated from longitudinal only initial conditions under
these fluid equations, as discussed in the next section, but the
transverse kinetic energy is still small enough in comparison to the
longitudinal kinetic energy for the above equation to be
approximately valid. After the shocks have formed, the energy
spectrum has the familiar behaviour of $k^{-2}$ in the inertial range.
According to Saffman, in the case of Burger's equation, the spectrum in the
inertial range has the form
\begin{equation} \label{saffman_spec}
E_\parallel(k) = \frac{\mathcal{L} \overline{J}^2}{4 \pi k^2} \, ,
\end{equation}
which we assume to also hold for the fluid equations employed in this
paper by applying the physical interpretations of $\mathcal{L}$ and
$\overline{J}^2$ to the longitudinal case. Here $\mathcal{L}$ is the mean
length of shocks per unit area, and $\overline{J}^2$ is the mean square jump
in velocity across the shock. In analogy to~\cite{saffman1971}, we cut off the
integral at the wavenumber corresponding to the length scale of the shock width
$\delta_s$ and substitute (\ref{saffman_spec}) into (\ref{viscousdiss}), which
after integration gives
\begin{equation} \label{KinEnDifEq}
\half \frac{d \left\langle \bv^2_\parallel \right\rangle}{dt} = - \frac{\mu}{2
\pi (1+c_s^2)} \frac{\mathcal{L} \overline{J}^2}{\delta_s} \propto
\frac{\bar{v}^3(t)}{L(t)} \, ,
\end{equation}
where to obtain the latter expression we have used the proportionality
relations 
\begin{equation} \label{PropRels}
\quad \mathcal{L} \propto L^{-1} \, , \quad \overline{J}^2 \propto \bar{v}^2 \,
,
\end{equation}
and the definition for the shock width $\delta_s$ in equation (\ref{swidth}).
Now in order to make progress, we need to find a relation between the time
behaviour of the integral length scale and the rms velocity. To this end, we
write the spectrum in the form
\begin{equation} \label{BrokPowLawk}
E_\parallel (k, t) = D(t) \frac{ \left[ k/k_p(t) \right]^\beta}{
1+ \left[ k/k_p(t) \right]^\alpha} \, ,
\end{equation} 
where the prefactor $D(t)$ contains the time dependence of the spectral
magnitude. Here we have ignored the high-$k$ behaviour of the spectrum found in
the previous section. Now in the low-$k$ power law range, when $k \ll k_p$, the
spectrum becomes
\begin{equation}
E_\parallel (k, t) \approx D(t) k_p(t)^{- \beta} k^\beta \, .
\end{equation}
The very low-$k$ end of the spectrum stays mostly unchanged, maintaining its
magnitude and power law index, as seen in Figure~\ref{fig:energyspecs}. Hence,
it can be approximated that
\begin{equation} \label{IRpowlaw}
D(t) k_p(t)^{- \beta} = \text{const.}
\end{equation}
Substituting this spectrum into Equation~(\ref{En_spec_def}) gives
\begin{equation}
\half \left\langle \bv_\parallel^2 \right\rangle = D(t) \int\limits_0^{\infty}
\frac{ \left[ k/k_p(t) \right]^\beta}{1+ \left[ k/k_p(t) \right]^\alpha} \,
dk \, .
\end{equation}
Using this form for the spectrum leads to an overestimation of the integral,
since we have ignored the high-$k$ behaviour, but we argue that this does not
affect the value of the energy significantly, since the largest contribution to
the integral comes from the energy containing scales around the peak of the
spectrum, and the contributions from scales smaller than the Taylor microscale
are small in comparison. After a change of variables $s = k/k_p$ the integral
becomes
\begin{equation}
\half \left\langle \bv_\parallel^2 \right\rangle = D(t) k_p(t)
\int\limits_0^\infty \frac{s^\beta}{1+s^\alpha} \, ds \, .
\end{equation}
Since the power laws stay the same after the shocks have formed, the parameters
$\alpha$ and $\beta$ are mostly constant. Thus, the integral gives
approximately a constant value, and by using the definition of the rms velocity
and relation (\ref{IRpowlaw}) with $k_p(t)^{-1}=L(t)$, we get
\begin{equation} \label{CondForL}
\bar{v}^2 (t) L(t)^{\beta + 1} = \text{const.} \equiv \xi^{-(\beta + 1)} \, .
\end{equation}  
Using this, we can now solve Equation~(\ref{KinEnDifEq}) for the energy
$\mathcal{E} = \left\langle \bv_\parallel^2 \right\rangle/2=\bar{v}^2/2$ with
the initial condition ${\mathcal{E}(t=0)=\mathcal{E}_0}$, yielding
\begin{equation} \label{KinEnAnalytical}
\mathcal{E}(t) = \frac{\mathcal{E}_0}{
  \left( 1 + C \frac{t}{t_s} \right)^{\zeta}} \, , \quad \zeta =
  \frac{2 (\beta + 1)}{\beta + 3} \, ,
\end{equation}
where in the denominator we have used (\ref{CondForL}) to write the constant 
$\xi$ in terms of the initial value of the integral length scale $L_0$ and the
initial energy $\mathcal{E}_0$, resulting in $\xi (2 \mathcal{E}_0)^{(\beta + 
3)/(2(\beta + 1))}=L_0^{-1} \bar{v}_0 = t_s^{-1}$, which is used as an estimate
for the shock formation time. We have also absorbed all constants into the
parameter $C$, whose value depends on the values of the prefactors of the
relations listed in Equation~(\ref{PropRels}). Without knowing the numerical
values of the prefactors, the value of $C$ can be obtained by fitting.
Based on our fits detailed later in this section, its typical value lies
in the range between 0.29 and 0.47. With the help of the above result,
Equation~(\ref{CondForL}) can be used to find $L(t)$, which reads as
\begin{equation} \label{IntLenAnalytical}
L(t) = L_0 \left( 1 + C \frac{t}{t_s} \right)^\lambda \, , \quad \lambda =
\frac{2}{\beta + 3} \, .
\end{equation}
The integral length scales of some of the runs featured in
table~\ref{tab:table3} of Appendix~\ref{AppB} are plotted in
Figure~\ref{fig:l_int_lon} against time in units of $t_s$ the shock formation
time.
\begin{figure}
\begin{center}
\includegraphics[width=\columnwidth]{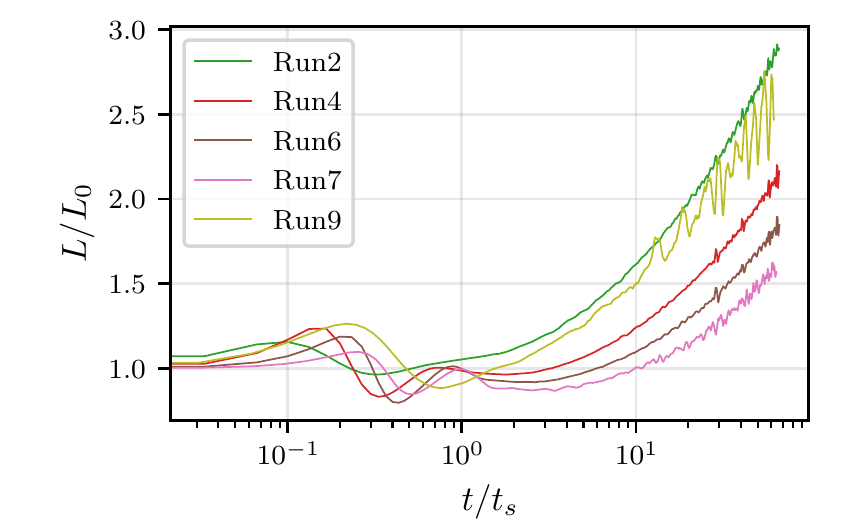}
\end{center}
\caption{\label{fig:l_int_lon} The longitudinal integral length scale
normalised by the initial value plotted for multiple runs as a function of the
number of shock formation times.}
\end{figure}
The results obtained for the power law values as functions of the
low-$k$ power law $\beta$ in equations (\ref{KinEnAnalytical}) and
(\ref{IntLenAnalytical}) coincide with those found by similar
  methods for three-dimensional classical vortical
turbulence~\cite{lesieur2008turbulence}. 

\begin{figure*}
\centering
\subfloat[]{
\includegraphics[width=0.5\textwidth]{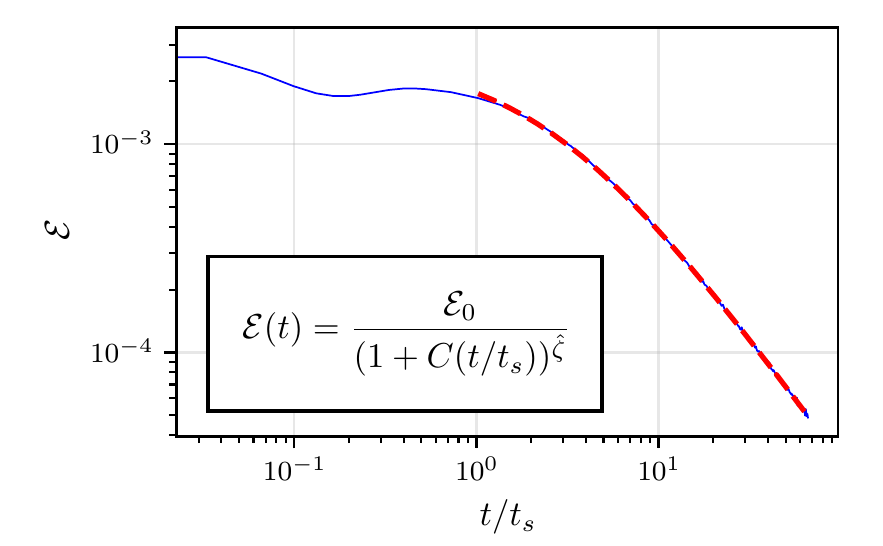}
\label{fig:kin_en_fit}
}
\subfloat[]{
\includegraphics[width=0.5\textwidth]{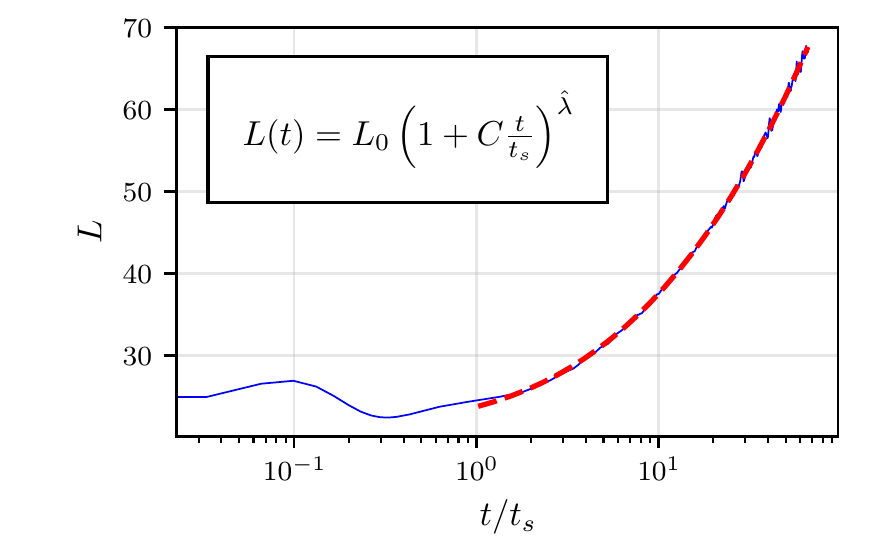}
\label{fig:l_int_fit}
}
\caption{Fits of the functions (\ref{KinEnAnalytical}) and
(\ref{IntLenAnalytical}) (red dashed curves) to the time evolution of the
kinetic energy (a) and the integral length scale (b) of run 2 (blue curves).
The fitting range is $t/t_s \geq 1$ and the parameter values obtained are
$\mathcal{E}_0 \approx 0.00263$, $\zeta \approx 1.207$, and $C \approx 0.391$
for the kinetic energy, and $L_0 \approx 20.8$, $\lambda \approx 0.346$ and $C
\approx 0.455$ for the integral length scale.}
\label{fig:kin_en_L_fits}
\end{figure*}

Equations (\ref{KinEnAnalytical}) and (\ref{IntLenAnalytical}) can now be used
as fitting functions to the curves seen in Figures~\ref{fig:kin_en}
and~\ref{fig:l_int_lon}, and the obtained power law indices can then be
compared to the analytical ones by measuring the low-$k$ power law index
$\beta$ of each run. The fits to the kinetic energy and the integral length
scale are constrained to the shock containing phase by using fitting ranges
whose lower boundary lies in the range $t/t_s \geq 1$. In these ranges the
inertial range has a power law of $k^{-2}$ and the fitting equations are valid.
Figure~\ref{fig:kin_en_L_fits} shows a pair of such fits for a single run. We
have varied the lower bound of the fit to all data points in the range
$1 \leq t/t_s \leq 3$ and averaged over the results to obtain the averaged
power law indices $\hat{\zeta}$ and $\hat{\lambda}$.
The low-$k$ power law indices are measured by fitting a broken power law, akin
to that in Equation~(\ref{BrokPowLawk}), on a suitable wavenumber range and
averaging the obtained values for the fit parameters over times $8 \leq t/t_s
\leq 12$, which in our simulations results to around 120 data points on
average. The suitable range in question has been chosen to be
$k \in [1/6L, 1/L_T]$, which contains the inertial range and a sufficient
amount of the low-$k$ power law. Time averaging like this is necessary because
there are oscillations in the spectrum that the fitting algorithm is sensitive
to. We have used the standard deviations of the time averaging to quantify the
strength of these oscillations.

It is also possible to derive relations between $\beta$, $\zeta$, and $\lambda$
that can be used to test the robustness of the theory by comparing to the
values obtained from the simulations by fitting. Such relations have been
obtained in Refs.~\cite{Olesen:1996ts, Brandenburg:2016odr} by considering
appropriate scaling of the energy spectrum and by making use of the rescaling
invariance of the hydrodynamic equations. Here, one relation follows
immediately from Equation~(\ref{CondForL}), which requires
\begin{equation} \label{ParamCond1}
\lambda (\beta + 1) - \zeta = 0
\end{equation}
for it to be valid. This relation can also be obtained directly from the power
laws in equations (\ref{KinEnAnalytical}) and (\ref{IntLenAnalytical}). A
relation containing only $\zeta$ and $\lambda$ can also be derived by replacing
$\beta$ in the equation above by using either of these two equations, giving
\begin{equation} \label{ParamCond2}
\zeta - 2(1-\lambda) = 0 \, .
\end{equation}

\begin{figure}
\begin{center}
\includegraphics[width=\columnwidth]{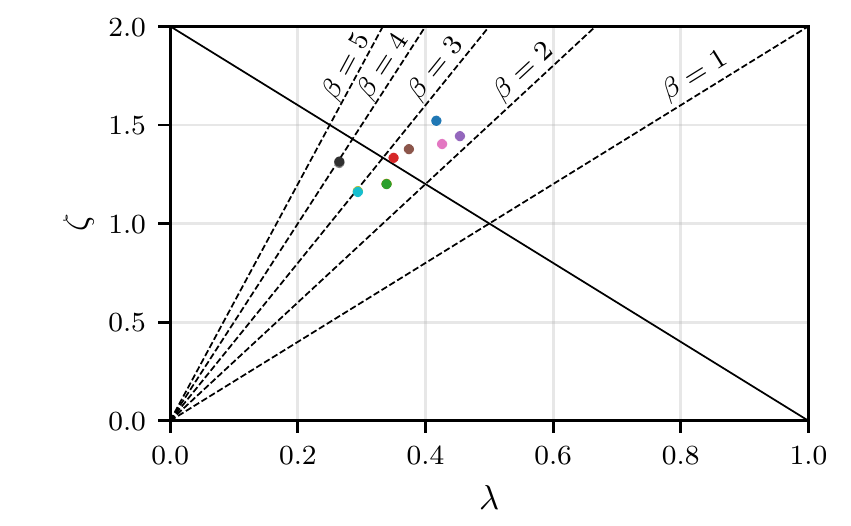}
\end{center}
\caption{\label{fig:ZLplot} A $\zeta \lambda$-plot that illustrates the
relations in equations (\ref{ParamCond1}) and (\ref{ParamCond2}). The diagonal
solid black line is the curve $\zeta = 2(1-\lambda)$. The data points for each
run have been obtained using the values of $\hat{\zeta}$ and $\hat{\lambda}$
from Table~\ref{tab:table1}.}
\end{figure}

Table~\ref{tab:table1} lists the averaged power law indices and the standard
deviations obtained from fits to the time evolutions of the kinetic energy and
the integral length scale. Power laws obtained from time averaging are denoted
by hats, and alongside them are the power laws obtained from equations
(\ref{KinEnAnalytical}) and (\ref{IntLenAnalytical}) using the values obtained
for the time averaged low-$k$ power law $\hat{\beta}$. These values are listed
in Table~\ref{tab:table2} alongside $\hat{\alpha}$ and the averaged value of
the inertial range power law $\widehat{\beta - \alpha}$. Also listed are the
standard deviations of these averages, denoted by sigmas, the initial low-$k$
power law index of the energy spectrum $\beta_0$ and the initial high-$k$ power
law $\beta_0 - \alpha_0$. The errors obtained from the fitting covariances are
negligible in comparison to the standard deviations of the time fluctuations in
all of these cases. We have also measured the magnitude of the statistical
fluctuations resulting from different initial random phases given to the
Fourier velocity components by making runs with the same initial conditions but
with different random seeds. Based on these runs, the fluctuations are found to
be either smaller or at the largest comparable in magnitude to the standard
deviations in Tables~\ref{tab:table1} and~\ref{tab:table2}. The values in these
two tables are used to test the relations (\ref{ParamCond1}) and
(\ref{ParamCond2}), which are listed in Table~\ref{tab:table2b} along with
their standard deviations obtained from the error propagation formula. These
are denoted as $\Delta C_i$ where the index $i$ marks the column of the table
(the run ID column being column 0). These relations are also plotted in a
$\zeta \lambda$-coordinate system in Figure~\ref{fig:ZLplot} where different
low-$k$ power law values correspond to lines with different slopes converging
at the origin~\cite{Brandenburg:2016odr}. The diagonal solid black line is the
curve $\zeta = 2(1-\lambda)$ of Equation~(\ref{ParamCond2}). The error bars for
the data points obtained from Table~\ref{tab:table1} are smaller than the data
point markers and are thus not drawn in the figure. The scaling law following
from the self-similarity is fulfilled well, with the value of zero lying within
the margin of error, whereas the one using the scaling invariance is not as
good due to the small standard deviations in the values of $\zeta$ and
$\lambda$.

\begin{table}
\begin{ruledtabular}
\begin{tabular}{D{.}{.}{1.0} D{.}{.}{1.3} D{.}{.}{1.3} D{.}{.}{1.3}
  D{.}{.}{1.3} D{.}{.}{10.0} D{.}{.}{10.0}}
\multicolumn{1}{c}{ID} &\multicolumn{1}{c}{$\hat{\zeta}$} &\multicolumn{1}{c}{
  $\zeta$} &\multicolumn{1}{c}{$\hat{\lambda}$} &\multicolumn{1}{c}{$\lambda$}
  &\multicolumn{1}{c}{$\sigma_\zeta$} &\multicolumn{1}{c}{$\sigma_\lambda$} \\
\hline
\rule{0pt}{3ex}
\textcolor{col1}{1} & 1.521 & 1.294 & 0.417 & 0.353 & $\num{1.09E-02}$
& $\num{7.48E-03}$ \\
\textcolor{col2}{2} & 1.201 & 1.252 & 0.339 & 0.374 & $\num{2.58E-03}$
& $\num{1.43E-03}$ \\
\textcolor{col3}{3} & 1.200 & 1.252 & 0.339 & 0.374 & $\num{2.56E-03}$
& $\num{1.42E-03}$ \\
\textcolor{col4}{4} & 1.333 & 1.284 & 0.350 & 0.358 & $\num{1.08E-02}$
& $\num{5.07E-03}$ \\
\textcolor{col5}{5} & 1.443 & 1.246 & 0.454 & 0.377 & $\num{1.85E-02}$
& $\num{1.48E-02}$ \\
\textcolor{col6}{6} & 1.377 & 1.359 & 0.374 & 0.320 & $\num{1.53E-02}$
& $\num{7.96E-03}$ \\
\textcolor{col7}{7} & 1.403 & 1.352 & 0.426 & 0.324 & $\num{1.83E-02}$
& $\num{1.32E-02}$ \\
\textcolor{col8}{8} & 1.307 & 1.330 & 0.265 & 0.335 & $\num{4.54E-03}$
& $\num{6.60E-03}$ \\
\textcolor{col9}{9} & 1.164 & 1.296 & 0.294 & 0.352 & $\num{3.07E-03}$
& $\num{2.07E-03}$ \\
\textcolor{col10}{10} & 1.160 & 1.237 & 0.294 & 0.381 & $\num{3.38E-03}$
& $\num{1.88E-03}$ \\
\textcolor{col11}{11} & 1.314 & 1.422 & 0.265 & 0.289 & $\num{7.46E-03}$
& $\num{4.15E-03}$
\end{tabular}
\end{ruledtabular}
\caption{\label{tab:table1}
Time averaged fit parameters for the kinetic energy and integral length scale
power laws $\hat{\zeta}$ and $\hat{\lambda}$, obtained by fitting the curves
seen in Figure~\ref{fig:kin_en_L_fits} so that the lower boundary of the
fitting range uses all data points in the range $1 \leq t/t_s \leq 3$, and by
averaging over the results. Also listed are the standard deviations, and the
predicted values for the power laws given by equations (\ref{KinEnAnalytical})
and (\ref{IntLenAnalytical}) by using the values for the time averaged low-$k$
power law $\hat{\beta}$ listed in Table~\ref{tab:table2}.}
\end{table}

\begin{table}
\begin{ruledtabular}
\begin{tabular}{D{.}{.}{1.0} D{.}{.}{1.0} D{.}{.}{3.0} D{.}{.}{1.3}
  D{.}{.}{1.3} D{.}{.}{2.3} D{.}{.}{1.3} D{.}{.}{1.3} D{.}{.}{1.3}}
\multicolumn{1}{c}{ID} &\multicolumn{1}{c}{$\beta_0$} &\multicolumn{1}{c}{
  $\beta_0 - \alpha_0$} &\multicolumn{1}{c}{$\hat{\alpha}$}
  &\multicolumn{1}{c}{$\hat{\beta}$}
  &\multicolumn{1}{c}{$\widehat{\beta-\alpha}$}
  &\multicolumn{1}{c}{$\sigma_\alpha$} &\multicolumn{1}{c}{$\sigma_\beta$}
  &\multicolumn{1}{c}{$\sigma_{\beta-\alpha}$} \\
\hline
\rule{0pt}{3ex}
\textcolor{col1}{1} & 4 & -8 & 3.576 & 2.669 & -0.907 & 0.368 & 0.167
& 0.363 \\
\textcolor{col2}{2} & 3 & -3 & 4.464 & 2.348 & -2.116 & 0.072 & 0.113
& 0.069 \\
\textcolor{col3}{3} & 3 & -3 & 4.464 & 2.349 & -2.115 & 0.072 & 0.113
& 0.068 \\
\textcolor{col4}{4} & 4 & -5 & 4.182 & 2.586 & -1.596 & 0.522 & 0.440
& 0.189 \\
\textcolor{col5}{5} & 5 & -15 & 4.238 & 2.305 & -1.933 & 0.426 & 0.299
& 0.182 \\
\textcolor{col6}{6} & 5 & -5 & 4.986 & 3.240 & -1.745 & 0.381 & 0.453
& 0.139 \\
\textcolor{col7}{7} & 9 & -6 & 5.258 & 3.176 & -2.082 & 0.261 & 0.264
& 0.027 \\
\textcolor{col8}{8} & 5 & -2 & 4.891 & 2.967 & -1.924 & 0.443 & 0.464
& 0.050 \\
\textcolor{col9}{9} & 3 & -3 & 4.693 & 2.683 & -2.010 & 0.710 & 0.739
& 0.052 \\
\textcolor{col10}{10} & 3 & -3 & 4.374 & 2.243 & -2.131 & 0.669 & 0.681
& 0.017 \\
\textcolor{col11}{11} & 7 & -4 & 5.970 & 3.916 & -2.055 & 0.514 & 0.530
& 0.026 
\end{tabular}
\end{ruledtabular}
\caption{\label{tab:table2}
The initial low-$k$ power law of energy spectrum $\beta_0$ and the initial
inertial range power law $\beta_0 - \alpha_0$, and the same parameters after
the shocks have formed obtained by time averaging the results obtained from
broken power law fits of Equation~(\ref{BrokPowLawk}) over the interval
$8 \leq t/t_s \leq 12$, denoted by hats. The last three columns list the
standard deviations for the time fluctuations of the parameters $\alpha$ and
$\beta$, and the inertial range power law.}
\end{table}

\begin{table}
\begin{ruledtabular}
\begin{tabular}{D{.}{.}{1.0} D{.}{.}{2.3} D{.}{.}{2.3} D{.}{.}{1.3}
  D{.}{.}{10.0}}
\multicolumn{1}{c}{ID} &\multicolumn{1}{c}{$\hat{\lambda} (\hat{\beta} + 1)
- \hat{\zeta}$} &\multicolumn{1}{c}{$\hat{\zeta} - 2(1-\hat{\lambda})$}
&\multicolumn{1}{c}{$\Delta C_1$} &\multicolumn{1}{c}{$\Delta C_2$} \\
\hline
\rule{0pt}{3ex}
\textcolor{col1}{1} & 0.008 & 0.354 & 0.076 & $\num{1.85E-02}$ \\
\textcolor{col2}{2} & -0.066 & -0.122 & 0.039 & $\num{3.85E-03}$ \\
\textcolor{col3}{3} & -0.066 & -0.122 & 0.039 & $\num{3.82E-03}$ \\
\textcolor{col4}{4} & -0.078 & 0.033 & 0.155 & $\num{1.48E-02}$ \\
\textcolor{col5}{5} & 0.056 & 0.351 & 0.145 & $\num{3.49E-02}$ \\
\textcolor{col6}{6} & 0.209 & 0.125 & 0.173 & $\num{2.21E-02}$ \\
\textcolor{col7}{7} & 0.376 & 0.255 & 0.126 & $\num{3.21E-02}$ \\
\textcolor{col8}{8} & -0.256 & -0.163 & 0.126 & $\num{1.40E-02}$ \\
\textcolor{col9}{9} & -0.080 & -0.247 & 0.218 & $\num{5.15E-03}$ \\
\textcolor{col10}{10} & -0.206 & -0.252 & 0.201 & $\num{5.06E-03}$ \\
\textcolor{col11}{11} & -0.013 & -0.157 & 0.142 & $\num{1.12E-02}$
\end{tabular}
\end{ruledtabular}
\caption{\label{tab:table2b} Numerical values for the relations of equations
(\ref{ParamCond1}) and (\ref{ParamCond2}) that are obtained using the fit
parameters in Tables~\ref{tab:table1} and~\ref{tab:table2}. The last two
columns contain the standard deviations of the relations in columns 1 and 2
obtained using the standard deviations of the fit parameters with the error
propagation formula.}
\end{table}

\subsection{Generation of transverse kinetic energy} \label{transenerg}
\noindent In our simulations we see an emergence of small amounts of transverse
kinetic energy from longitudinal-only initial conditions. In order to study the
vorticity generation more closely, we can take a look at the vorticity
equation, obtained by taking a curl of Equation~(\ref{EulerEq}). The equation
can be written for the vorticity $\omega = \nabla \times \bv$, which in
two-dimensional case can be treated as a scalar, giving
\begin{align}
\frac{\partial \omega}{\partial t} + (1-2c_s^2) \omega (\nabla \cdot \bv)
&+ (1-c_s^2) (\bv \cdot \nabla) \omega \\ \nonumber
&- c_s^2 \bv \times \nabla^2 \bv = \frac{\eta}{1+c_s^2} \nabla^2 \omega \, .
\end{align}
From this it follows that if initially $\omega = 0$
\begin{equation}
\frac{\partial \omega}{\partial t} = c_s^2 \bv \times
\nabla ( \nabla \cdot \bv) \, ,
\end{equation}
meaning that there is a vorticity generating term resulting from the last term
on the left hand side of (\ref{EulerEq}), giving rise to some transverse
kinetic energy even when the initial conditions contain only longitudinal
modes.

\begin{figure*}
\centering
\subfloat[$v(x, y) = |\mathbf{v}|$]{
\includegraphics[width=0.5\linewidth]{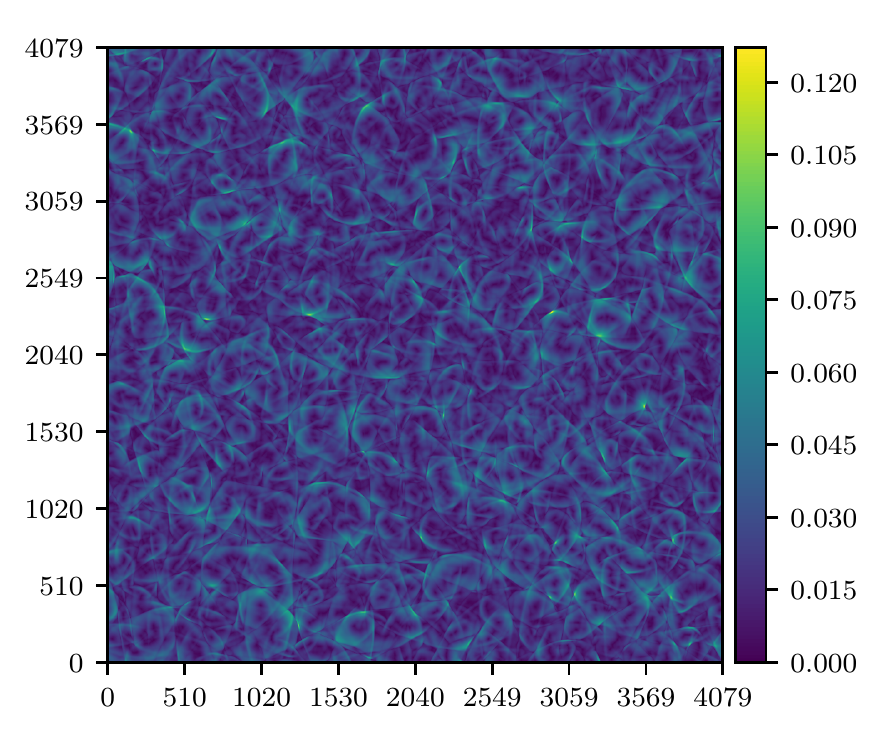}
\label{fig:vmag1}
}
\subfloat[$L \omega (x, y)$]{
\includegraphics[width=0.5\linewidth]{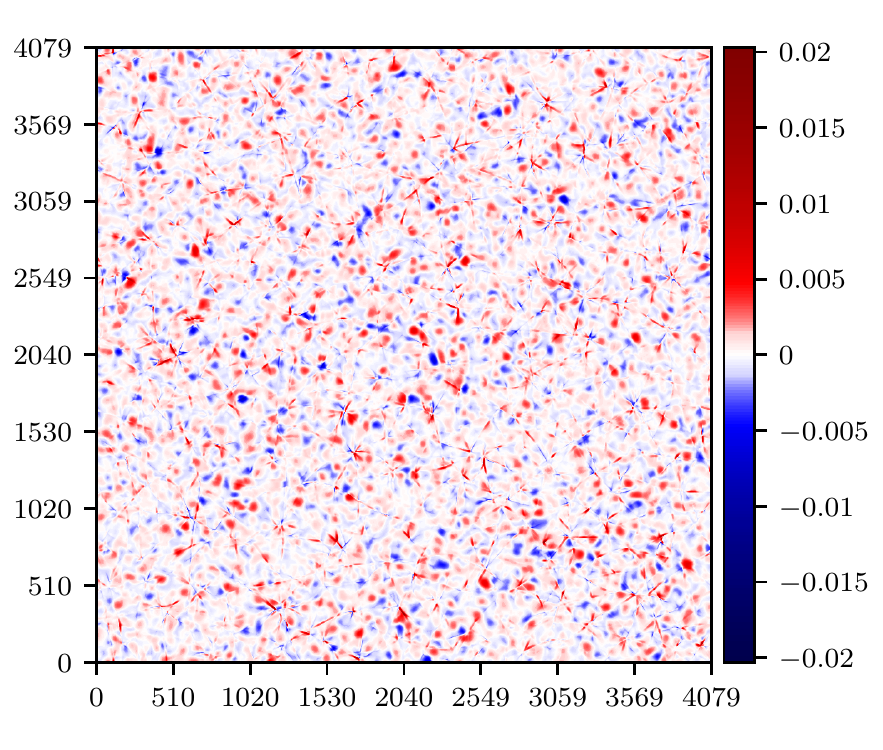}
\label{fig:vor1}
} \\
\subfloat[$v(x, y) = |\mathbf{v}|$]{
\includegraphics[width=0.5\linewidth]{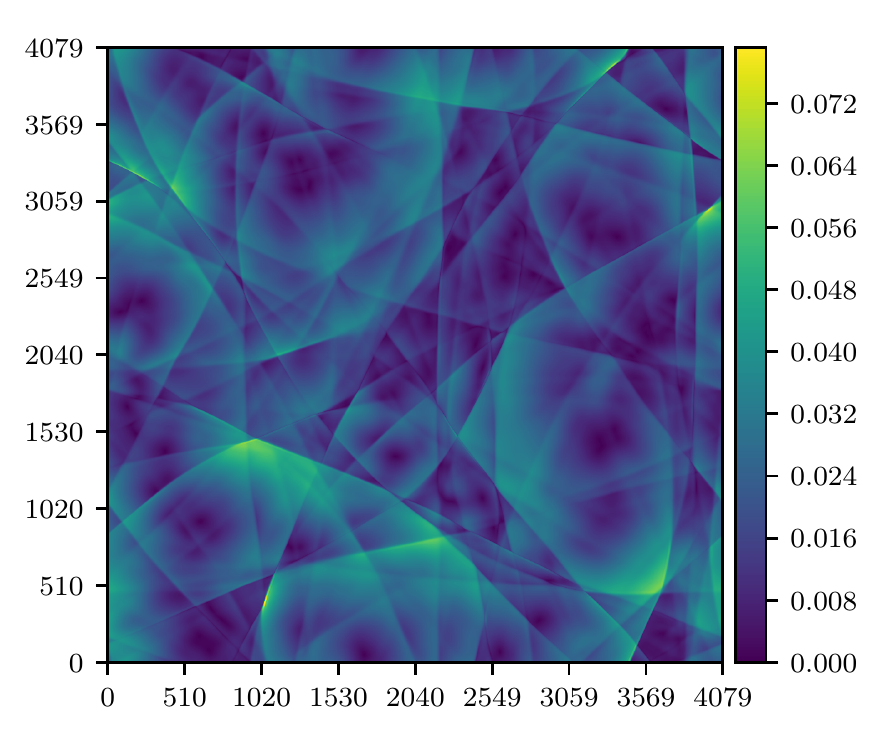}
\label{fig:vmag2}
}
\subfloat[$L \omega (x, y)$]{
\includegraphics[width=0.5\linewidth]{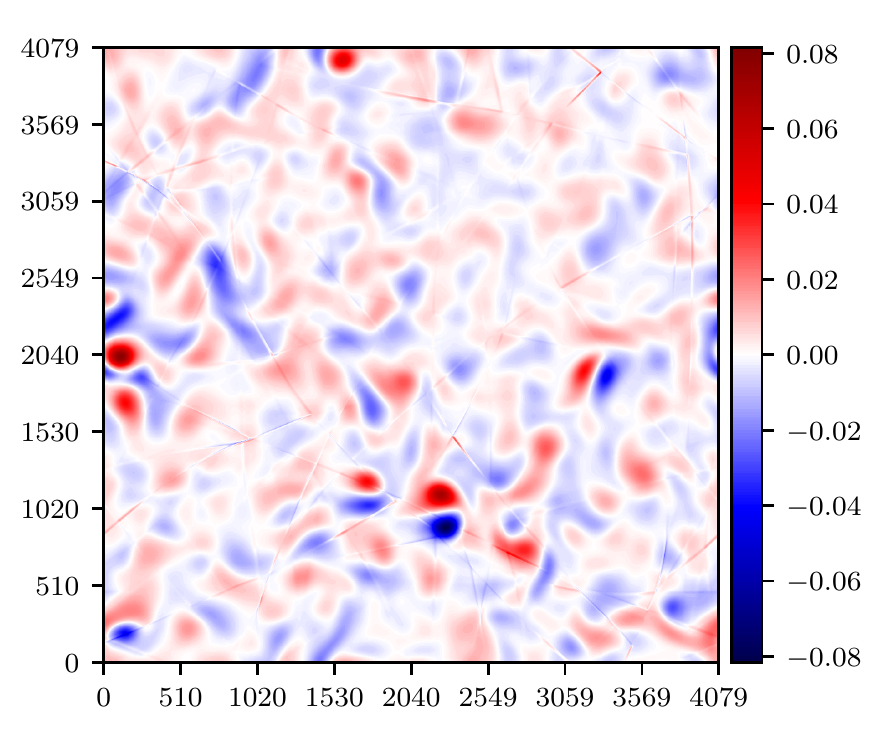}
\label{fig:vor2}
}
\caption{The magnitude of the velocity field $|\bv|$ (a) and the corresponding
vorticity field $\omega = \nabla \times \mathbf{v}$ scaled by the integral
length scale (b) of a moderate Reynolds number $4080^2$ resolution run after
about 13 shock formation times. Figures (c) and (d) show the same quantities
for a high Reynolds number run at the end of the run at about $t = 67 t_s$. The
runs have the same initial conditions as runs 2 and 9. In Figures (b) and (d)
vortex-like structures can be seen, appearing in pairs of different signs.}
\label{fig:vmagvor}
\end{figure*}

\begin{figure}
\begin{center}
\includegraphics[width=\columnwidth]{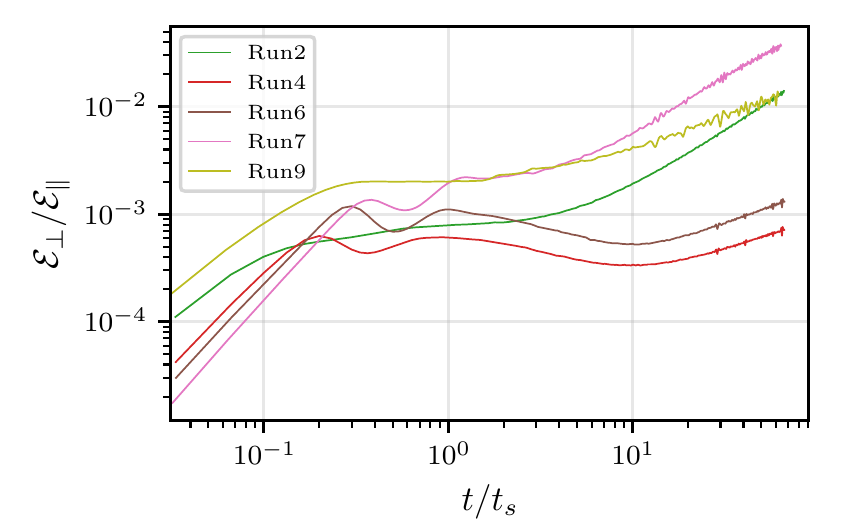}
\end{center}
\caption{\label{fig:lontranen}
Development of the ratio of the transverse to longitudinal kinetic energy of
several runs with time, in units of the shock formation time $t_s$. The high
Reynolds number runs 7 and 9, and the low Reynolds number runs 2
and 4 are clearly separated into two groups by at least an order of
magnitude, with the high Re runs having higher transverse kinetic energy. The
exception is the low Re bulk viscosity only run (ID 2, green), which joins the
high Re curves at the end of the run.}
\end{figure}

In the simulations we see that early on the vorticity field attains its largest
values in the regions containing overlapping or colliding shocks. This is
illustrated in Figures~\ref{fig:vmag1} and~\ref{fig:vor1} that show the
magnitude of the velocity field and the corresponding vorticity that has been
scaled by the integral length scale to obtain a dimensionless quantity. As the
shocks overlap with each other, their amplitude increases, and regions with
much higher amplitudes than seen in the initial conditions are formed, shown in
the figure in yellow. The largest values of vorticity right after the shocks
are formed are obtained around these regions, shown as thin short dark red
lines in the contour plot. These features are short-lived and change location
as the shocks travel. The other part of the vorticity field after shock
formation is the background vorticity that changes slowly in comparison to the
vorticity from shock collisions, and contains vortex-like structures that often
appear in pairs of different signs. Over time as the shocks get dissipated, the
background vorticity becomes dominant, with the shocks being only faintly
visible in comparison, as seen in Figure~\ref{fig:vor2}, which plots the
dimensionless vorticity field at the very end of a run. The higher the Reynolds
number of the run is, the higher the generated transverse kinetic energy is
relative to the longitudinal kinetic energy. In Figure~\ref{fig:lontranen} the
energy fraction $\mathcal{E}_\perp/ \mathcal{E}_\parallel$ has been plotted for
several runs, and the group of curves with the highest values corresponds to
the high Reynolds number runs. It shows that the transverse kinetic energy is
still small compared to the longitudinal kinetic energy, even after 60 shock
formation times.

Runs 3 and 10 contain only bulk viscosity. We find that in the longitudinal
case, both the bulk and the shear viscosity affect the fluid almost in an
identical way. Runs 2 and 3, and 9 and 10 have the same initial conditions and
random phases, with the only difference being the viscosity type. The values
for the viscosities in these runs are chosen so that the value of the effective
viscosity is the same. In the longitudinal case, these pairs of runs produce
results that are very close to each other, which can also be seen from the
plots of longitudinal quantities, such as in
Figures~\ref{fig:kin_enA},~\ref{fig:l_int_lonA} of Appendix~\ref{AppC},
and~\ref{fig:ZLplot}, where the runs overlap, or from the tables of the
previous section. 

The same is not true in the transverse case, as is evident by
Figures~\ref{fig:lontranen} and \ref{fig:lontranenA}, where the curves of the
previously mentioned run pairs clearly separate from each other some time after
the start of the run, with the bulk viscosity only run having a larger
transverse kinetic energy at the end in both cases (see Table \ref{tab:table3}
in appendix \ref{AppB} for the colour and viscosity type of each run in the
case of Figure~\ref{fig:lontranenA}). This is because in the shear viscosity
only case the dissipation of energy is larger, as under these fluid equations
the dissipation due to viscosity can be shown to be
\begin{equation}
\half \frac{d \left\langle \bv^2 \right\rangle}{dt} =
\begin{cases}
 - \frac{2 \mu}{1+\cs^2} \int\limits_0^\infty k^2 E(k) \, dk \, ,
 \quad \text{when } \nabla \times \bv=0  \\
 - \frac{2 \eta}{1+\cs^2} \int\limits_0^\infty k^2 E(k) \, dk \, ,
 \quad \text{when } \nabla \cdot \bv=0
\end{cases}
\end{equation}
meaning that for the transverse component the viscous dissipation is caused
only by the shear viscosity. This also strongly affects the shape of the
transverse energy spectrum at large-$k$ between the bulk and shear-viscosity
only runs. 

The focus of this paper is the study of the longitudinal case, and thus there
is more potential work to be done in studying the transverse case under these
fluid equations. The transverse only case (incompressible flow) has been
extensively studied and forms part of the standard understanding of turbulence
presented in textbooks (see e.g. Ref.~\cite{lesieur2008turbulence}).

\section{Estimate for the gravitational wave power spectrum} \label{GravSpec}
\noindent While there are no gravitational waves in two dimensions, we can
estimate the gravitational wave power spectrum generated by shocks in three
dimensions by using the results in Ref.~\cite{KP}, according to which the
energy spectrum maintains the $k^{-2}$ inertial range power law in any number
of space dimensions. By assuming the energy spectrum to have a simple broken
power law form, the GW power spectrum can be obtained by adapting standard
methods~\cite{Kosowsky:2001xp,Gogoberidze:2007an,Caprini:2007xq,Caprini:2009yp,
Caprini:2009fx,Hindmarsh:2019phv}.

The source of the gravitational waves is taken to be the shear stresses
resulting from a velocity field consisting of randomly distributed sound waves,
generated on a timescale long compared the light-crossing time of any important
scales in the velocity field. The resulting GW power spectrum can be calculated
from the unequal time velocity field correlators for the system. Our
calculation assumes that the shock lifetime $t_s$ is much less than a Hubble
time, meaning that the expansion of the universe can be approximated by
setting the velocities to zero after a Hubble time~\cite{Hindmarsh:2015qta}.
It is also assumed that the fluid velocities are non-relativistic, and that the
velocity can be treated as a Gaussian random field, with any non-Gaussianity
leading to negligible contributions to the connected four-point correlator. As
the initial velocity field steepens into shocks, the velocity field loses its
Gaussianity but we assume the deviation from Gaussianity to be small, so that
the correlator can still be approximately treated as Gaussian. Measuring the
unequal time correlators for a collection of shock waves to test the validity
of this assumption stands as possible future work.

We begin by citing Equation (3.46) of Ref.~\cite{Hindmarsh:2019phv}, which 
gives the growth rate of the gravitational wave power spectrum
$ \mathcal{P}_\text{gw}$ as
\begin{equation} \label{dPgwdt}
\frac{1}{H_*} \frac{d}{dt} \mathcal{P}_{\text{gw}} = 3
\left( \Gamma \bar{v}^2 \right)^2 (H_* L) \frac{(kL)^3}{2 \pi^2}
\tilde{P}_{\text{gw}} (kL) \, ,
\end{equation}
where $H_*$ is the Hubble rate at the time of the transition and $\Gamma$ is
the mean adiabatic index of the fluid. We take $\Gamma = (1+c_s^2) = 4/3$, as
appropriate for an ultrarelativistic fluid. The final factor in the expression
is a dimensionless spectral density function, defined as
\begin{align} \label{DSpecDen}
\tilde{P}_{\text{gw}} (y) = \frac{1}{4 \pi y c_s} &
\left( \frac{1-c_s^2}{c_s^2} \right)^2 \int\limits_{z_-}^{z_+} \frac{dz}{z}
\frac{(z-z_+)^2 (z-z_-)^2}{z_+ + z_- - z} \\ \nonumber
& \times \tilde{P}_v (z) \tilde{P}_v (z_+ + z_- - z) \, ,
\end{align}
where $z_\pm = y(1 \pm c_s)/(2 c_s)$, $z=qL$, and $\tilde{P}_v (z)$ is the
scaled velocity spectral density, which is related to the actual spectral
density as
\begin{equation} \label{scaledPv}
P_v(qL) = L^3 \bar{v}^2 \tilde{P}_{v} (qL)
\end{equation}
with $q$ being the wavenumber. The relation between the energy spectrum and
the spectral density in 3D is
\begin{equation} \label{3DPowSpecEnSpecRel}
E(k) = \frac{k^2}{2 \pi^2} P_v (k) \, .
\end{equation}
On the other hand, the energy spectrum can also be written in terms of the
collapse function $\Psi$ as seen in Equation~(\ref{EnPsi}), from which it
follows that
\begin{equation}
P_v (z) = \pi^2 L^3 \bar{v}^2 \frac{\Psi(z)}{z^2} \, .
\end{equation}
Now for the function $\Psi(z)$ we use the broken power law form of
Equation~(\ref{BrokPowLaw}) that by using Equation~(\ref{scaledPv}) gives
\begin{equation}
\tilde{P}_v (z) = \frac{\Psi_0 \pi^2}{z_p^2}
\frac{(z/z_p)^{\beta-2}}{1+(z/z_p)^\alpha} \, .
\end{equation}
Here the parameter $\kappa_p$ of Equation~(\ref{BrokPowLaw}) has been denoted
with $z_p$ to coincide notationally with $z$ and is fixed in terms of $\alpha$
and $\beta$ along with $\Psi_0$ through Equations (\ref{Afix}) and
(\ref{KappaFix}). Using this and Equation~(\ref{DSpecDen}), and integrating
Equation~(\ref{dPgwdt}) with respect to time with a change of variable
${z=kLs}$ gives the following expression for the gravitational wave power
spectrum
\begin{align} \label{PgwFin}
\frac{1}{(H_* L_0)^2} & \mathcal{P}_{\text{gw}} (k, t_{H_\star}) = \frac{3 \pi
\Psi_0^2 \Gamma^2 (1-c_s^2)^2}{8 L_0^2 z_p^4 c_s^5} k^{5} \\ \nonumber
& \times \int\limits_{0}^{t_{H_\star}} dt \, \bar{v}^4(t)  L^{6}(t)
\int\limits_{s_-}^{s_+} ds \, I(s, t)
\end{align}
where $t_{H_\star}$ is the lifetime of the GW source, which we recall is
taken to be the Hubble time at the time of the
phase transition,~\cite{Hindmarsh:2015qta}. The integrand $I$ has the form
\begin{equation}
I(s, t) = \frac{(s-s_+)^2 (s-s_-)^2 [s(s_+ + s_- - s)/s_p^2(t)]^{\beta - 3}}{
  s_p^2(t)[1+[s/s_p(t)]^\alpha][1+[(s_+ + s_- - s)/s_p(t)]^\alpha]} \, ,
\end{equation}
with $s_p(t) = z_p/kL(t)$, and $s_\pm = (1 \pm c_s)/(2 c_s)$\
Now we write the time integral only in terms of the integral scale $L(t)$ by
relating it to $\bar{v}(t)$ using equation (\ref{CondForL}) and
substituting the time development Equations (\ref{KinEnAnalytical}) and
(\ref{IntLenAnalytical}) into it (while keeping in mind that
$\mathcal{E} = \bar{v}^2/2$). Because of the relation between the power
law indices in Equation~(\ref{ParamCond1}) the time dependence vanishes and the
equation can be written in the form
\begin{equation}
\bar{v}^2 (t) = \bar{v}_0^2 \left( \frac{L(t)}{L_0} \right)^{-(\beta + 1)} \, .
\end{equation}
Here $\bar{v}_0$ denotes the initial value of the rms velocity. Using
this, the time integral in Equation~(\ref{PgwFin}) can be written as
\begin{equation} \label{TimeInt}
  \int\limits_{0}^{t_{H_\star}} dt \, \bar{v}^4(t)  L^{6}(t) = 
  \bar{v}_0^4 L_0^6 \int\limits_{0}^{t_{H_\star}} dt \,
  \left( \frac{L(t)}{L_0} \right)^{2(2-\beta)} \, .
\end{equation}
Next we make a change of variables
$\tau = k L(t) / z_p = s_p^{-1}$ in the time integral. This is tantamount to
integrating over the integral length scale, which in the scenario considered
here is a monotonically increasing quantity with time.
The time differential can be related to the differential of this new
variable by using Equation~(\ref{IntLenAnalytical}), which yields
\begin{equation}
  dt = \frac{z_p}{\lambda C k \bar{v}_0}
  \left( \frac{\tau}{\tau_0} \right)^{1 / \lambda-1} \, d \tau \, ,
\end{equation}
where $\lambda$ is the decay power law of the integral length scale, $C$ is a
decay parameter whose inverse gives the number of shock formation times that it
takes for the flow to start decaying, and $\tau_0 = k L_0 / z_p$. Now, the $s$
integrand can also be written as
\begin{equation}
  I(s, \tau) = \tau^{2(\beta - 2)} I_\tau (s, \tau) \, ,
\end{equation}
where
\begin{equation}
  I_\tau (s, \tau) = \frac{(s-s_+)^2 (s-s_-)^2 \left[ s (s_+ + s_- - s)
\right]^{\beta-3}}{\left[ 1 + ( \tau s)^\alpha \right] \left[ 1 +
\tau^\alpha (s_+ + s_- - s)^\alpha \right]} \, ,
\end{equation}
meaning that the factor of $\tau$ resulting from Equation~(\ref{TimeInt}) ends
up cancelling with that coming from the $s$ integrand. The
Equation~(\ref{PgwFin}) now becomes
\begin{align}
  &\frac{1}{(H_* L_0)^2} \mathcal{P}_{\text{gw}} (k) = \frac{3 \pi
\Psi_0^2 z_p \Gamma^2 (1-c_s^2)^2}{8 c_s^5} \\ \nonumber
& \times \left( \frac{k L_0}{z_p} \right)^4 \tau_0^{2 \beta - 1/\lambda - 3}
\int\limits_{\tau_0}^{\tau_{H_\star}} d \tau \, \tau^{1/\lambda - 1}
\int\limits_{s_-}^{s_+} ds \, I_\tau(s, \tau) \, ,
\end{align}
where $\tau_{H_\star} = k L(t_{H_\star}) / z_p$, and in the prefactor the
powers of $k$ and $L_0$ are equal, so that they can be written in terms of
$\tau_0 = k L_0 / z_p$. It is worth noting that with this formulation the
integration limits also depend on the wavenumber $k$. We can now write the
power law index $\lambda$ in terms of the low-$k$ power law index of the energy
spectrum $\beta$ using the relation between them in
Equation~(\ref{IntLenAnalytical}). We then factorise the result and write it in
the form
\begin{equation} \label{GWspecTau}
\frac{1}{(H_* L_0)^2} \mathcal{P}_{\text{gw}} (k L_0, \tau_{H_\star})
= \frac{\bar{v}_0^3}{C} \mathcal{N} S(k L_0, \tau_{H_\star}) \, ,
\end{equation} 
where the numerical factor $\mathcal{N}$ is determined by the speed of sound in
the fluid and the power law parameters $\alpha$ and $\beta$ appearing in the
energy spectrum and has the form
\begin{equation}
\mathcal{N} = \frac{3 \pi}{8} \frac{(\beta + 3) \Psi_0^2 z_p}{2}
\frac{\Gamma^2 (1-c_s^2)^2}{c_s^5} \, .
\end{equation}
The function $S(k L_0, \tau_{H_\star})$ determines the shape of the
spectrum, and can be written as
\begin{equation} \label{GWSpecShape}
S(k L_0, \tau_{H_\star}) = \tau_0^{(3 \beta -1)/2}
\int\limits_{\tau_0}^{\tau_{H_\star}} d \tau \,
\tau^{(\beta + 1)/2} \int\limits_{s_-}^{s_+} ds \, I_\tau (s, \tau) \, .
\end{equation}

\begin{figure}
\begin{center}
\includegraphics[width=\columnwidth]{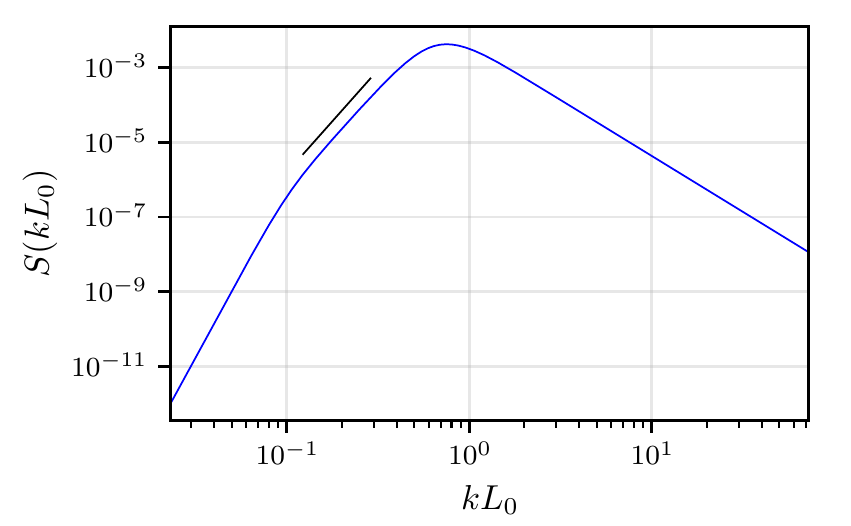}
\end{center}
\caption{\label{fig:GWspec} The function $S(k L_0)$ obtained numerically from
Equation (\ref{GWSpecShape}) with parameter values $\alpha = 6$, $\beta = 4$,
and $L(t_{H_\star})/L_0 = 6.1$. A bend in the spectrum at the low-$k$ end is
seen at $k L_0 \simeq z_p L_0/L(t_{H_\star}) \simeq 0.1$. The black line
demonstrates a power law of $k^{5.5}$.}
\end{figure}

Using Equation (\ref{GWSpecShape}), we have plotted the shape of the GW power
spectrum numerically. In the three-dimensional case the low-$k$ power law
index of the energy spectrum $\beta$ is not expected to be the same as in 2D,
and should be determined by numerical simulations. After the
phase transition has completed, the fluid contains shocks and has the $k^{-2}$
power law at the inertial range. For this estimate, we have assumed a value of
$\beta = 4$, and taken $\alpha = 6$ to obtain the correct value for the
high-$k$ power law. The spectrum is then obtained by numerically integrating
the two integrals that appear in (\ref{GWSpecShape}) for a given ratio
$L(t_{H_\star})/L_0$, which we have taken to be 6.1 for
illustrative purposes when plotting the spectrum in Figure~\ref{fig:GWspec}.
The figure highlights an
interesting aspect in the low-$k$ end of the spectrum, in that there is a
change in the low-$k$ power law index around
$k L_0 \simeq z_p L_0/L(t_{H_\star}) \simeq 0.095$, after which the power law
changes from a steeper $k^{9}$ power law to a shallower power law of $k^{5.5}$.
The location where this change occurs is determined by the lifetime of the
source $t_{H_\star}$ through the ratio $L(t_{H_\star})/L_0$, so that the
shallower power law appears in the range
\begin{equation}
z_p L_0/L(t_{H_\star}) \lesssim kL_0 \lesssim z_p \, .
\end{equation}
Therefore, for short enough lifetimes, where the integral scale does not have
enough time to grow significantly compared to its initial value, the range is
short and close to the peak, meaning that effectively only the steeper slope is
obtained, and for long lifetimes, where $L(t_{H_\star}) >> L_0$, the bend
occurs at very small wavenumbers close to the origin, so that the spectrum
effectively only possesses the shallower slope. In the first case, the shock
formation time $t_s$ is close in magnitude to the duration of the GW source
$t_{H_\star}$, which is the Hubble time.  Hence only short-lived source
$t_s \ll t_{H_\star}$, as assumed here, will show the intermediate power law.

\begin{figure}
\begin{center}
\includegraphics[width=\columnwidth]{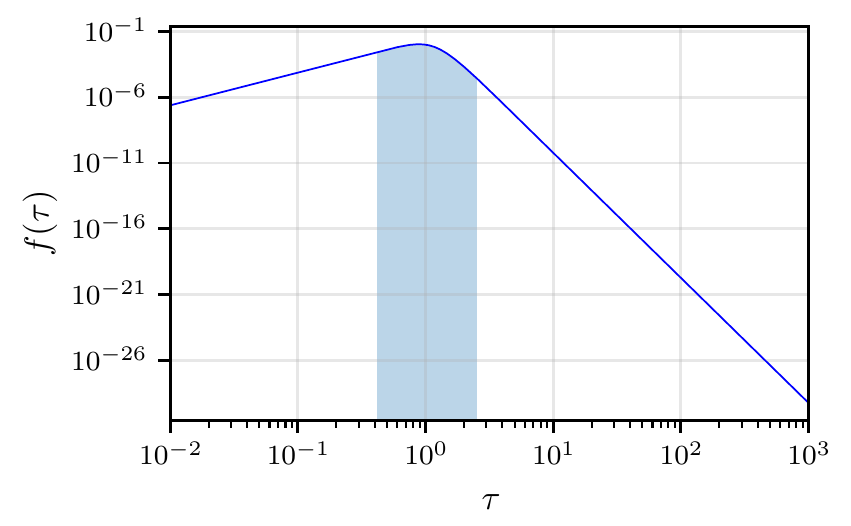}
\end{center}
\caption{\label{fig:ftau} The integrand function $f(\tau)$
(see Eq.~\ref{taufunc}) plotted using the same parameter values as in
Figure~\ref{fig:GWspec}. The highlighted area shows the part of the curve
contributing to the GW power spectrum at $k L_0=0.24$, which lies roughly in
the middle of the intermediate power law range.}
\end{figure}

The power law behaviour of the GW power spectrum can be inspected by extracting
the wavenumber behaviour of Equation (\ref{GWspecTau}) in different limits. At
very small wavenumbers fulfilling the condition $\tau \ll 1$ for any $\tau \in
[\tau_0, \tau_{H_\star}]$ the integral over $s$ yields essentially a constant,
from which it follows that
\begin{equation}
\frac{1}{(H_* L_0)^2} \mathcal{P}_{\text{gw}} (k) \propto k^{2 \beta + 1} \, ,
\end{equation}
which for $\beta = 4$ gives the value of the power law index seen in
Figure~\ref{fig:GWspec}. At large wavenumbers, so that $\tau \gg 1$ for any
$\tau \in [\tau_0, \tau_{H_\star}]$, it can be approximated
\begin{equation}
I_\tau (s, \tau) \approx \frac{(s-s_+)^2 (s-s_-)^2 \left[ s (s_+ + s_- - s)
\right]^{\beta-\alpha-3}}{\tau^{2 \alpha}} \, ,
\end{equation}
which means that the $s$-integral yields a constant once more, and after
integrating over $\tau$, the $k$-dependence is found to be
\begin{equation}
\frac{1}{(H_* L_0)^2} \mathcal{P}_{\text{gw}} (k) \propto k^{2(\beta-\alpha)+1}
\, ,
\end{equation}
which for acoustic turbulence gives the power law of $k^{-3}$ at high
wavenumbers. To touch on the intermediate power law seen in
Figure~\ref{fig:GWspec}, we need to understand the behaviour of the
$\tau$-integrand in the regime where the $s$-integral does not yield a
constant. To this end, we rewrite the integrals of Equation~(\ref{GWSpecShape})
in the form
\begin{equation}
\frac{1}{(H_* L_0)^2} \mathcal{P}_{\text{gw}} (k)
\propto k^{\frac{3 \beta - 1}{2}} \int\limits_{\tau_0}^{\tau_{H_\star}}
f(\tau) \, ,
\end{equation}
where the function $f(\tau)$ denotes the integrand
\begin{equation} \label{taufunc}
f(\tau) = \tau^{\frac{\beta+1}{2}} \int\limits_{s_-}^{s_+} ds \,
I_\tau (s, \tau) \, .
\end{equation}
This function has been obtained numerically by using the same parameter values
as in Figure~\ref{fig:GWspec}, and is plotted in Figure~\ref{fig:ftau}.
Since the integration limits depend on the wavenumber $k$, a different part of
this curve is integrated for each value of $k$. It turns out that the
intermediate power law is obtained at wavenumbers for which the integration
range spans the peak of the function $f(\tau)$, that is, when the separation
between $\tau_{H_\star}$ and $\tau_0$ is larger than the width of the peak in
the integrand $f(\tau)$, which is located approximatively in the range
$0.5 \lesssim \tau \lesssim 2$. The width of the integration range for a given
$k$ is determined by the the ratio $L(t_{H_\star})/L_0$. When it is large, the
peak is panned even for small wavenumbers, resulting in the narrower power law
at low-$k$, and when it is small, the integration range is narrow and does not
span the peak entirely for any $k$ so that only the steeper power law is
obtained. For the wavenumbers in the intermediate power law range, the integral
over $f(\tau)$ is effectively a constant, since the largest contribution to
the integral is obtained around the peak, which is spanned for all such
wavenumbers, and since the contributions from the edges of the integration
range are small in comparison. Therefore, it follows that in the intermediate
power law range the GW power spectrum goes as
\begin{equation}
\frac{1}{(H_* L_0)^2} \mathcal{P}_{\text{gw}} (k)
\propto k^{\frac{3 \beta - 1}{2}} \, ,
\end{equation}
giving the power law seen in Figure~\ref{fig:GWspec} when $\beta = 4$.

To conclude, apart from giving a power law of $k^{-3}$ in the high-$k$ range,
the decay of the shocks also induces a change in the low-$k$ power law, going
from $k^{2 \beta + 1}$ to a shallower $k^{(3 \beta - 1)/2}$ one, over a range
depending on the integral scale of the fluid flow after a Hubble time. Note
that the rate at which the flow was originally generated may also appear as a
scale in the gravitational wave power spectrum, below which another power law
may apply~\cite{Caprini:2009fx}. We have assumed that this happens at a lower
wavenumber than any considered here.

\section{Conclusions}

\noindent We have studied decaying acoustic turbulence using two-dimensional
numerical simulations with the emphasis being on the impact of the shocks upon
the energy spectrum, and on the decay of the kinetic energy. Conducting the
simulations in two dimensions allows for better computational efficiency and
the use of larger grid sizes in comparison to 3D, which leads to there being
more dynamic range in the wavenumber space. Two-dimensional systems are also
simpler to analyse and in the case of shocks share some properties with
three-dimensional systems. By making use of the universality of the power
spectra, the obtained two-dimensional decay properties and power laws of the
system have been applied in three dimensions to calculate an estimate for the
gravitational wave power spectrum resulting from a collection of shock waves.

The longitudinal energy spectrum of the fluid can be written in terms of the
longitudinal kinetic energy, integral scale, and the dimensionless function
$\Psi(k L)$ as seen in Equation~(\ref{EnPsi}). The function $\Psi(kL)$ has the
property that it maintains its shape over time at length scales above the
dissipation range. Using the tanh shock profile obtained from the fluid
equations, we have presented an analytical universal form for this function,
which is found to be a broken power law modulated by an integral function
$\mathcal{I}$ that is shown in Equation~(\ref{IP}). This function depends on
the steepness of the shocks via the wavenumber parameter $k_s$ appearing in the
argument of the tanh shocks. Between the wavenumbers corresponding to the
integral scale and the Taylor microscale, the power law is found to be
$k^{-2.08 \pm 0.08}$, which agrees very well with the $k^{-2}$ power law
associated with acoustic turbulence~\cite{KP}, obtained as an inverse-variance
weighted average of the measurements in Table \ref{tab:table2}. At lower
wavenumbers, using the same method, the power law is $k^\beta$, with
$\beta = 2.50 \pm 0.31$.

In order to find the time evolution of the longitudinal kinetic energy, we have
used the $k^{-2}$ inertial range power law, and the self-similarity of the
spectrum at low-$k$ to find equations (\ref{KinEnAnalytical}) and
(\ref{IntLenAnalytical}), the latter of which describes the decay of the
longitudinal integral length scale. At times much larger than the shock
formation time, these produce power law forms, where the values of the power
law indices depend on the low-$k$ power law index of the energy spectrum. From
the simulations using the earlier averaging technique with the means and
standard deviations listed in Table \ref{tab:table1}, we find the kinetic
energy to decay as $t^{-1.21 \pm 0.06}$, and the integral scale to increase as
$t^{0.32 \pm 0.03}$. To test the validity of our results, we have used the
analytical results and the scaling relations between the power law parameters,
and compared the results from those to the independent data obtained from the
simulations by fitting. In general, we find these to be in good agreement.

Lastly, we have produced an estimate for the shape of the gravitational wave
power spectrum in three dimensions, using the universality of the $k^{-2}$
spectrum for a shocked fluid, and the evolution laws for the kinetic energy and
the integral scale. The power spectrum is peaked at a wavenumber set by the
initial integral scale. At higher wavenumbers the GW power spectrum is found to
go as $k^{-3}$, which is the same as the power law predicted from linear
evolution of acoustic waves produced by first order phase transitions~\cite{
  Hindmarsh:2017gnf,Hindmarsh:2019phv}.

At wavenumbers lower than the peak of the spectrum, there is a change in the
power law from $k^{2 \beta + 1}$ to a less steep $k^{(3 \beta - 1)/2}$. This
power law is maintained down to values of $k$ of order the inverse integral
scale at the end of the effective sourcing of GWs, expected to be about a
Hubble time.

Our work is of direct relevance for calculations of the gravitational wave
power spectrum produced by first order thermal phase transitions in the early
Universe, in cases where the shock formation and decay time $t_s$ is shorter
than the Hubble time, often the case for phase transitions strong enough to be
observed.  The acoustic turbulence simulated here in two dimensions will also
develop in three dimensions, with the same $k^{-2}$ power law in the energy
spectrum at high $k$.  This is also the same power law as found in the linear
approximation to the evolution of the sound waves following the phase
transition, and so we do not expect qualitative changes to the GW power
spectrum as a result of the appearance and decay of shocks. However, we do
expect the acoustic turbulence to significantly affect the power-law behaviour
of the gravitational wave power spectrum at wavenumbers lower than the peak,
where a non-trivial power law may develop in the energy spectrum. The index of
this power law cannot, however, be predicted from two-dimensional numerical
simulations.  In any case, the low-$k$ power law in the gravitational wave
spectrum will be different from that from the linear evolution of acoustic
waves and from vortical turbulence.  Finding this characteristic power law is
clearly a high priority for reliable predictions for the gravitational wave
power spectrum following a phase transition.

\begin{acknowledgments}
\noindent 
We acknowledge useful discussions with Carl Bender. J.D was supported by the
Magnus Ehrnrooth Foundation. D.J.W. (ORCID ID 0000-0001-6986-0517) was
supported by Academy of Finland grant nos. 324882 and 328958, J.D. and K.R.
(ORCID ID 0000-0003-2266-4716) by Academy of Finland grant nos. 319066 and
320123, and M.~H. (ORCID ID 0000-0002-9307-437X) by Academy of Finland grant
no.~333609. The authors would also like to thank Finnish Grid and Cloud
Infrastructure at the University of Helsinki
(urn:nbn:fi:research-infras-2016072533) and CSC – IT Center for Science,
Finland, for computational resources.
  
\end{acknowledgments}

\appendix

\section{Shock tube runs} \label{AppA}
\noindent In the appendices, the length and time units are the lattice spacing.
We take the speed of light to be 1. In order to check the validity of the
results obtained for the shock waves in Section~\ref{shockshape}, we have
conducted runs on a shock tube, a very thin lattice of size $12240 \times 2$,
essentially corresponding to a one-dimensional situation. The grid spacing and
the time step size are still the same as before. The initial condition in the
energy density is a waveform
\begin{multline}
\delta \rho(x) = \frac{1}{2} \left[ \tanh \left( \frac{1}{4}
\left( x + \frac{N}{20} \right) \right) \right. \\
\left. - \tanh \left( \frac{1}{4}
\left( x - \frac{N}{20} \right) \right) \right] \, ,
\end{multline}
where $N=12240$, giving a nearly square shaped waveform whose center is
located at the origin and whose width is about $10 \%$ of the grid length. The
velocity is zero initially, that is $v_x (\mathbf{x}, \mathbf{y}) = v_y
(\mathbf{x}, \mathbf{y}) = 0$. These initial conditions do not fulfil the
requirement $\left\langle \delta \rho  \right\rangle = 0$, but this is not
essential, since we are only interested in the shocks and their properties, and
not in the physicality of the system. The initial waveform breaks into two
shocks, one travelling to the right and one to the left, with such waves also
appearing in the velocity.

\begin{figure}
  \centering
  \subfloat[Velocity]{
    \includegraphics[width=\columnwidth]{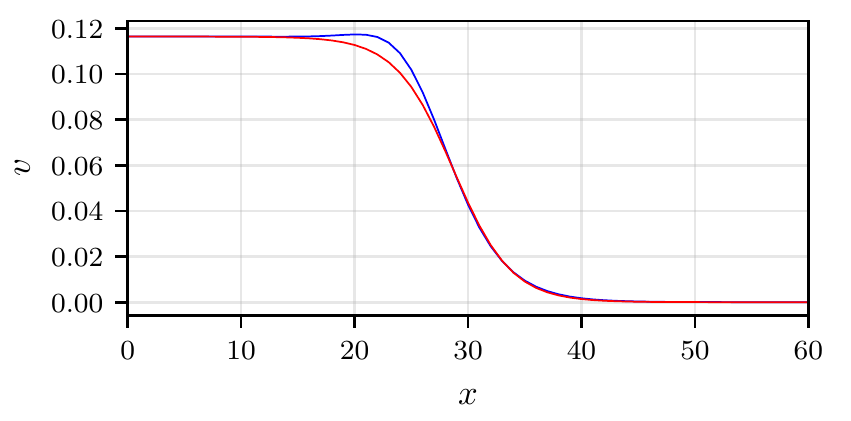}
    \label{fig:shockcompA}
  } \\
  \subfloat[Energy density]{
    \includegraphics[width=\columnwidth]{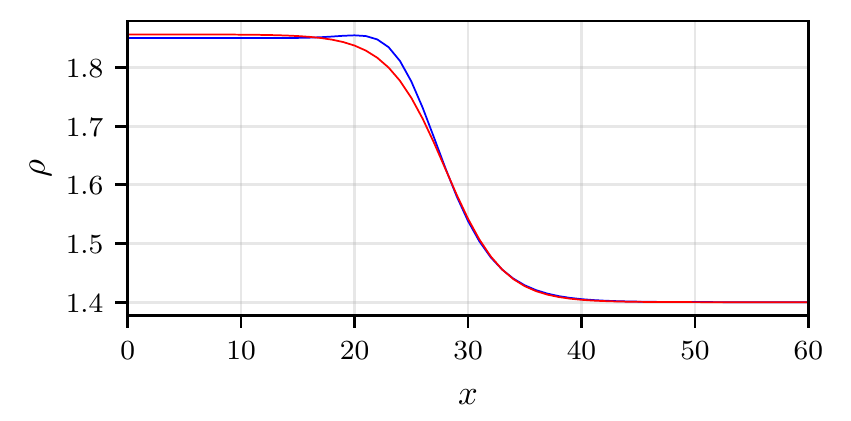}
    \label{fig:shockcompB}
  }
  \caption{\label{fig:shockcomp} Comparison between a right-moving shock in
  (a) velocity and (b) energy density obtained from simulation data
  (blue line) and the shock profile predicted by section~\ref{shockshape}
  (red line) early on in a shock tube run. The viscosity of the run is
  $\eta = 0.264$, and the parameter values used for the shock profile are
  $V_+ = 0.116$, $V_m = 0$, and $x_0 = 28.7$, giving a shock velocity
  $u \approx 0.632$.}
\end{figure}

\begin{figure}
\begin{center}
\includegraphics[width=\columnwidth]{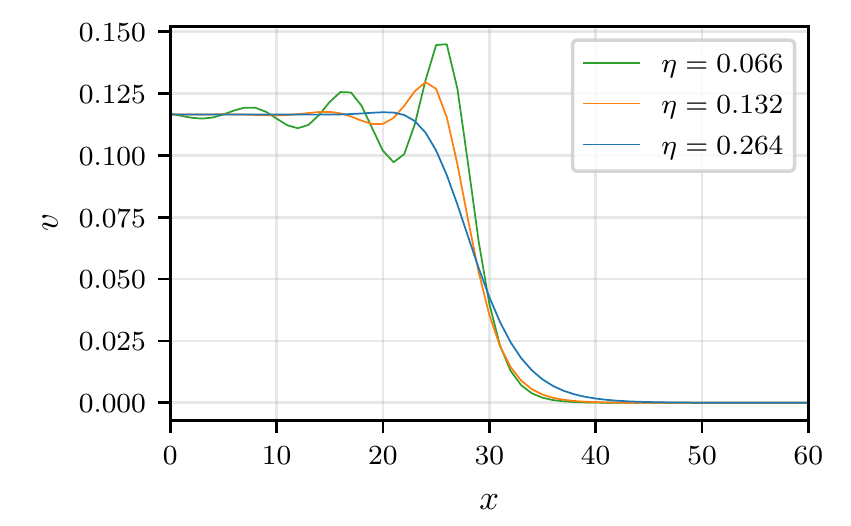}
\end{center}
\caption{\label{fig:shockcomp2} Shock profiles of different shock tube runs
with varying values of shear viscosity after 6000 simulation time units.}
\end{figure}

\begin{figure}
\begin{center}
\includegraphics[width=\columnwidth]{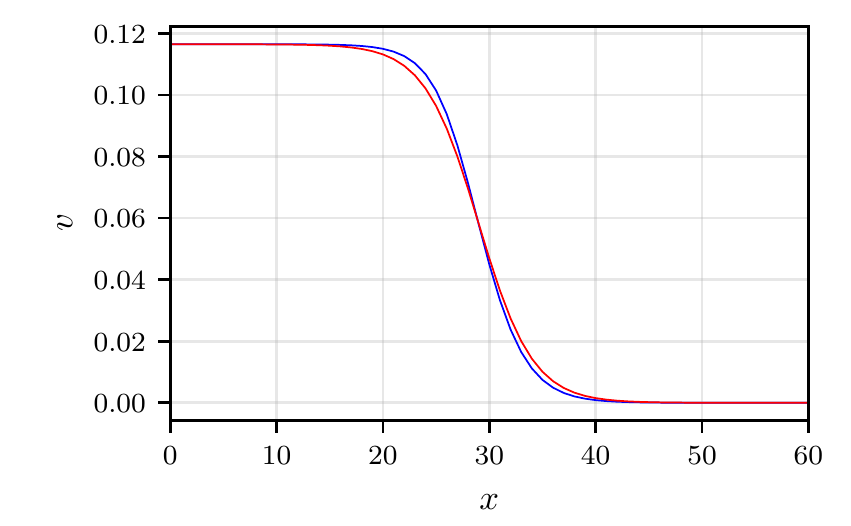}
\end{center}
\caption{\label{fig:shockcomp4} Figure~\ref{fig:shockcompA} but a fourth order
accurate finite difference scheme has been used in the run, improving the
correspondence of the curves. All parameter values are the same apart from
$x_0 = 29.0$.}
\end{figure}

These isolated shock waves can now be used to test the shock profile found in
Equation~(\ref{ShockProf}) by comparing it to the simulation data. This is done
in Figure~\ref{fig:shockcompA} for the velocity, which shows the wave profile
obtained from data in blue and the shock profile obtained from the
aforementioned equation in red for a run with a shear viscosity value of
$\eta = 0.264$. The same is done for the energy density in
Figure~\ref{fig:shockcompB} where to obtain the red curve the relation
\begin{equation}
  \delta \rho (x, t) = \rho_0 \left[ \frac{u}{u - (1+c_s^2) V(x,t)} - 1 \right]
\end{equation}
is used, which reduces to Equation~(\ref{VelDenRel}) when
$\delta \rho \ll \rho_0$.
We see that the model matches the data quite well, apart from the crest of the
shock, wherein there is a slight deviation from the value predicted by the
model caused by the numerical scheme's inability to precisely deal with sharp
discontinuities. The strength of this effect depends on the initial rms
velocity of the run, and the value of the viscosity, with large rms velocities
and small viscosities leading to larger deviations and more oscillatory
behaviour. This is demonstrated in Figure~\ref{fig:shockcomp2} where a
right-moving shock profile has been plotted for runs with varying viscosity
values. This effect can be reduced by using a higher order finite difference
scheme, as can be seen in Figure~\ref{fig:shockcomp4}, where a fourth order
scheme has been used in performing the same run as in
Figure~\ref{fig:shockcompA}. By impacting the shock shape, higher order schemes
also slightly change the shape of the energy spectrum at highwavenumbers and
also, based on our tests, increase the amount of transverse power generated.
Neither of these have a significant impact on the key results we have
presented\footnote{The change in the high-$k$ end of the spectrum resulting
from the use of a higher order finite difference scheme changes the value
obtained for the parameter $\kappa_s$ of Figure~\ref{fig:Psifit} in
Section~\ref{EnSpec} but the fit is still good.}.
Another aspect to consider is the conservation of the energy density
$\left\langle \rho \right\rangle = \rho_0$, which is not by default taken into
account when a central difference scheme is used. We have measured how well
this conservation is fulfilled and we find that the largest deviations are
obtained in the high Reynolds number runs. In run 9 the largest deviation for
$\left\langle \rho \right\rangle / \rho_0$ from unity is 0.1, which is obtained
only briefly at the start of the run, when the shocks are at their strongest,
after about 4 shock formation times. The quantity remains mostly within
2.5\% of the expected value and the deviations from it show a decreasing trend
throughout the run after the initial phase. As these are the most extreme
deviations and considering their small magnitude, we conclude that the energy
density is conserved to a satisfactory level of accuracy even in the high
Reynolds number case and thus the central difference scheme we have employed is
expected to provide representative results for the set of runs we have featured
in this paper.

\begin{figure}
\begin{center}
\includegraphics[width=\columnwidth]{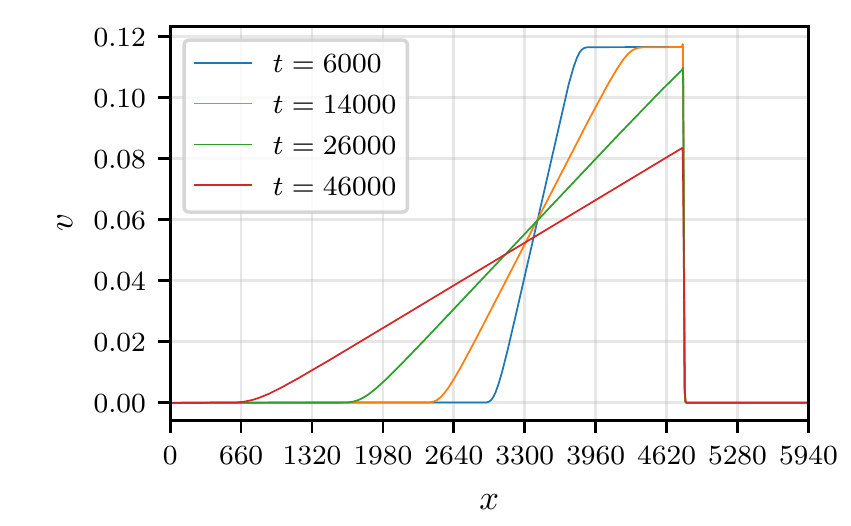}
\end{center}
\caption{\label{fig:shockcomp3} Evolution of the wave profile of a right-moving
shock in the $\eta=0.264$ run. The profiles at various times have been shifted
in $x$ to make the shocks overlap.}
\end{figure}

The time evolution of the wave profile in a $\eta = 0.264$ run has been plotted
in Figure~\ref{fig:shockcomp3}.
Over time the top of the wave profile gets narrower until a sawtooth form is
reached. There is no decrease in the amplitude of the shock before this point.
The opposite is true for the bottom of the wave profile, where it gets wider as
time goes on. Using these properties, we have measured the velocity of the
shock wave in this run and compared it to the value given by
Equation~(\ref{ShockVel}) for the shock wave seen in the simulation. The chosen
time window spans 6000 simulation time units, and is chosen near the start of
the run in such a way that the wave profiles are not yet in the sawtooth phase,
and there are no collisions between the right and left moving shock waves that
could affect the shock speed. Since there is no decay in the amplitude and no
change in the steepness of the shock, the propagation of a single point in the
waveform can be measured in this interval, assuming a constant velocity. The
value obtained from Equation~(\ref{ShockVel}) is found to be within 1\% of th
 measured value.

\section{Runs and initial conditions} \label{AppB}

\begin{table*}[t!]
\begin{ruledtabular}
\begin{tabular}{D{.}{.}{2.0} D{.}{.}{2.0} D{.}{.}{1.0} D{.}{.}{1.0}
  D{.}{.}{2.0} D{.}{.}{1.3} D{.}{.}{5.0} D{.}{.}{1.2} D{.}{.}{1.3}
  D{.}{.}{1.3} D{.}{.}{1.2} D{.}{.}{3.0} D{.}{.}{1.4} D{.}{.}{3.1}}
\multicolumn{1}{c}{$\text{ID}$} &\multicolumn{1}{c}{$\alpha_0$}
&\multicolumn{1}{c}{$\beta_0$} &\multicolumn{1}{c}{$\gamma$}
&\multicolumn{1}{c}{$\beta_0 - \alpha_0 + 1$} &\multicolumn{1}{c}{$k_p$}
&\multicolumn{1}{c}{$A/V$} &\multicolumn{1}{c}{$B$} &\multicolumn{1}{c}{$\eta$}
&\multicolumn{1}{c}{$\nu$} &\multicolumn{1}{c}{$|\delta|_{\text{max}}$}
&\multicolumn{1}{c}{$t_s$} &\multicolumn{1}{c}{$\bar{v}$}
&\multicolumn{1}{c}{$\text{Re}$} \\
\hline
\textcolor{col1}{1} & 12 & 3 & 4 & -8  & 0.100 & 64 & 0.5A & 0.066 & 0 & 0.12
& 166 & 0.0761 & 10.9 \\
\textcolor{col2}{2} & 6 & 2 & 3 & -3  & 0.035 & 128 & 1.0A & 0.066 & 0 & 0.66
& 302 & 0.0768 & 20.2 \\
\textcolor{col3}{3} & 6 & 2 & 3 & -3  & 0.035 & 128 & 1.0A & 0 & 0.088 & 0.66
& 302 & 0.0768 & 20.2 \\
\textcolor{col4}{4} & 9 & 3 & 3 & -5  & 0.035 & 352 & 0.5A & 0.066 & 0 & 0.14
& 301 & 0.0955 & 31.1 \\
\textcolor{col5}{5} & 20 & 4 & 5 & -15  & 0.035 & 900 & 0.5A & 0.066 & 0 & 0.12
& 532 & 0.0731 & 32.3 \\
\textcolor{col6}{6} & 10 & 4 & 2 & -5  & 0.025 & 800 & 0.5A & 0.066 & 0 & 0.08
& 300 & 0.1275 & 55.4 \\
\textcolor{col7}{7} & 15 & 8 & 3 & -6  & 0.009 & 17920 & 1.0A & 0.066 & 0
& 0.50 & 625 & 0.1519 & 164.1 \\
\textcolor{col8}{8} & 7 & 4 & 2 & -2  & 0.007 & 7840 & 1.0A & 0.066 & 0 & 0.44
& 322 & 0.2201 & 177.4 \\
\textcolor{col9}{9} & 6 & 2 & 3 & -3  & 0.007 & 15680 & 1.0A & 0.066 & 0 & 0.27
& 643 & 0.1726 & 217.5 \\
\textcolor{col10}{10} & 6 & 2 & 3 & -3  & 0.007 & 15680 & 1.0A & 0 & 0.088
& 0.27 & 643 & 0.1726 & 217.5 \\
\textcolor{col11}{11} & 11 & 6 & 5 & -4  & 0.006 & 89600 & 1.0A & 0.066 & 0
& 0.61 & 595 & 0.1830 & 224.2
\end{tabular}
\end{ruledtabular}
\caption{\label{tab:table3} The initial parameter values for the spectral
densities and some other quantities of interest in the $10080^2$-resolution
runs used in this paper.}
\end{table*}

\noindent The runs have been performed using code written in Python with
Cython~\cite{cython} providing C-like performance in the most computationally
demanding parts of the simulations, like in the evaluation of the spatial
derivatives over arrays. The code is parallelised using MPI for
Python~\cite{mpi4py} so that the computations can be distributed to multiple
processor cores to provide further speed ups. NumPy~\cite{2020NumPy-Array} has
been used for the computations involving arrays along with numexpr, which
accelerates computations between arrays and optimises memory usage. The Runs
have been conducted on CSC's (Finnish IT center for science) supercomputer
Puhti. All fits and numerical integrations used to obtain the results featured
in this paper have been performed with SciPy~\cite{2020SciPy-NMeth}, which is a
Python library offering tools for scientific computing. The routines used are
$\mathtt{curve\_fit}$ found in $\mathtt{scipy.optimize}$ for the fits, and
$\mathtt{quad}$ found in $\mathtt{scipy.integrate}$ for the numerical
integration. Numerical integration via $\mathtt{quad}$ is not however used in
calculating quantities whose definitions contain integrals over the energy
spectrum, such as the rms-velocities or the integral length scale in
Equation~(\ref{int_len_scale}). Instead, in those cases the integrals are
discretised as sums over the squared Fourier arrays as
\begin{equation}
\int d^2 k \rightarrow \frac{(2 \pi)^2}{V} \sum\limits_{\bar{k}} \, ,
\end{equation}
where $V=N^2 (\Delta x_1)(\Delta x_2)$. The routine from SciPy has only been
used in the evaluation of the integrals in equations (\ref{IP}), (\ref{PgwFin})
and (\ref{taufunc}).

The initial conditions are given in terms of the longitudinal and transverse
spectral densities
\begin{align}
P_\parallel (|\mathbf{k}|) &= \frac{1}{V} \Big( |v_x^\parallel (\mathbf{k})|^2 + |v_y^\parallel (\mathbf{k})|^2 \Big) \\
P_\perp (|\mathbf{k}|) &= \frac{1}{V} \Big( |v_x^\perp (\mathbf{k})|^2 + |v_y^\perp (\mathbf{k})|^2 \Big)
\end{align}
given in the form of Equation~(\ref{initPowSpec}). The real space velocity
components are then solved from these using the Fourier space projectors
\begin{align}
v_i^\perp (\mathbf{k}) &= (\delta_{ij} - \hat{k}_i \hat{k}_j)
v_j (\mathbf{k}) \\
v_i^\parallel (\mathbf{k}) &= \hat{k}_i \hat{k}_k v_j (\mathbf{k})
\end{align}
and by taking the inverse Fourier transforms of $v_i (\mathbf{k})$. Here
$\delta_{ij}$ is the Kronecker delta, and the Einstein summation convention is
applied. On the lattice, the unit vectors $\hat{k}_i$ are written in terms of
the eigenvalues of the derivative operators as
\begin{equation}
\hat{k}_i = \frac{\sin (k_i \Delta x_i)}{\sqrt{\sum\limits_i \sin^2
(k_i \Delta x_i)}} \, ,
\end{equation} 
where $\Delta x_i$ is the lattice spacing in the direction of the $i$:th
component. These projectors are also used to get the longitudinal energy
spectra at non-initial times by applying them to the Fourier transformed
velocity components. The Fourier transform algorithm utilised is the
N-dimensional Fast Fourier Transform routine provided by NumPy. The spectrum
$E_{\perp}(k)$ or $E_\parallel (k)$ is then obtained from the $N \times N$
sized arrays by radially averaging over circular rings of width $\Delta k$,
which is the reciprocal lattice spacing. The averaging stops when the edges of
the array are reached, meaning that the corner regions are ignored.

This paper contains results from 11 runs with a resolution of $10080^2$ whose
initial conditions are listed in Table~\ref{tab:table3}. In addition, there are
a couple of $4080^2$ resolution runs that are used in plotting the contour
plots of Figures~\ref{fig:rhodiv} and~\ref{fig:vmagvor} that use the same
initial conditions as runs 2 and 9. The runs listed in the table are labelled
from 1 to 11 in the order of increasing longitudinal Reynolds number. Columns
2-4 and 6-7 contain the initial parameter values for the longitudinal velocity
spectral density given in Equation~(\ref{initPowSpec}). The parameter $A$ has
been scaled by dividing it by the volume $V$ to reduce its magnitude.
The energy density is initialised in the same way as the velocity, and
the initial density spectral density in all of the runs is the same, apart from
the prefactor $A$ that is replaced by $B$, found in the 8th column of the
table, given in terms of $A$. The 5th column shows the value of the initial
high-$k$ power law in the energy spectrum.

The next two columns after the spectral parameters list the values of the shear
viscosity $\eta$ and the bulk viscosity $\nu$. In the runs featured in this
paper, all runs with shear viscosity use the value of $0.066$ and all runs with
bulk viscosity use the value $0.088$. Finding a suitable value for the
viscosity is a balancing act, as too large values either lead to no shocks
forming at all, or to the formation of very weak and short lived shocks,
whereas too low values give rise to significant instabilities and undesirable
effects like the appearance of strong oscillations at the crest of the shocks.
The bulk viscosity runs 3 and 10 use the same random seeds as runs 2 and 10,
meaning they have the same initial waveforms both in velocity and density. All
other runs use seeds that differ from each other.

The final four columns list some initial quantities measured from the initial
conditions. The quantity $|\delta|_\text{max}$ is the largest value obtained by
the fractional density perturbation $\delta \rho / \rho_0$ at the initial time,
and $t_s$ is the shock formation time. The length of the runs in simulation
time units is determined by it, as all of the runs are cut off after about 60
shock formation times. The final two columns list the initial root mean square
velocity, and the longitudinal Reynolds number of the run, obtained using
Equation~(\ref{ReynoldsNum}). The cutoff parameter $k_d$ seen in the
initial spectrum of equation~\ref{initPowSpec} has the value of $1/\sqrt{5}$
in all runs.

The simulation code that has been used to create these runs can be found in
Ref.~\cite{dahl_jani_2021_5786090}. Also included are the scripts used to
initialise each of these runs. Non-related movies of longitudinal and
transverse only runs produced with the simulation code can be found in
Refs.~\cite{dahl_jani_2021_5786442, dahl_jani_2021_5786393}.

\clearpage

\section{Plots containing all of the runs} \label{AppC}

\noindent This section contains versions of Figures~\ref{fig:kin_en},
\ref{fig:l_int_lon}, and \ref{fig:lontranen} where curves from all runs found
in table~\ref{tab:table3} are included in the plots. The runs are
distinguishable from each other by the colours found in the ID column of the
table.

\begin{figure}[h]
  \begin{center}
  \includegraphics[width=\columnwidth]{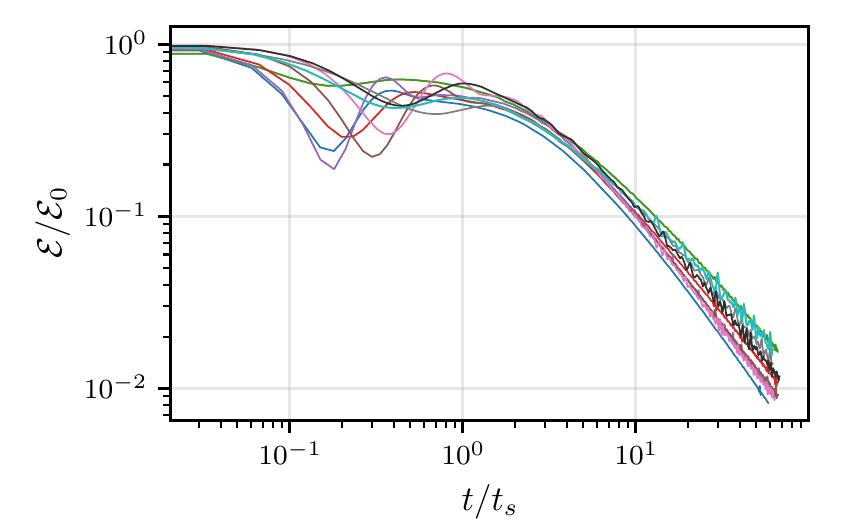}
  \end{center}
  \caption{\label{fig:kin_enA} Figure~\ref{fig:kin_en} but all runs featured in
    table~\ref{tab:table3} are included. The runs have been colour coded to
    match the colours found in the ID columns of the tables here and in all
    other figures containing multiple runs to allow for distinguishability.}
  \end{figure}

\begin{figure}[h]
  \begin{center}
  \includegraphics[width=\columnwidth]{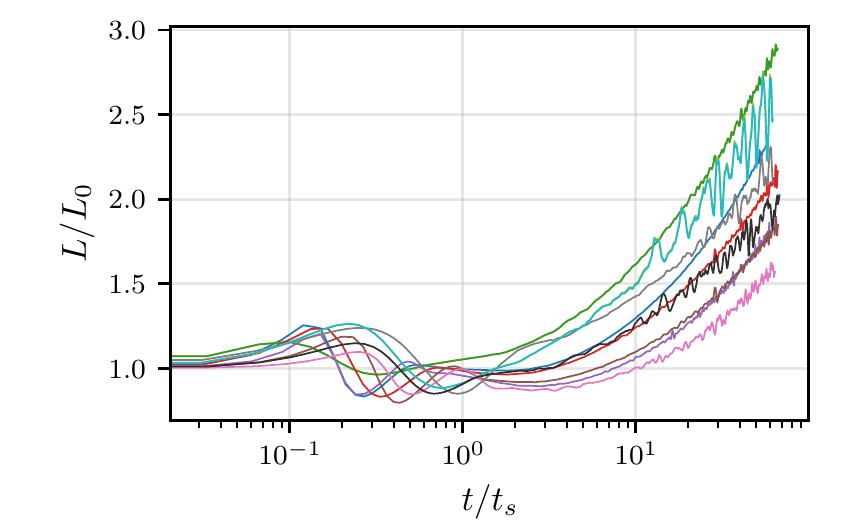}
  \end{center}
  \caption{\label{fig:l_int_lonA} Figure~\ref{fig:l_int_lon} but all runs
  featured in table~\ref{tab:table3} are included.}
  \end{figure}

  \begin{figure}[h]
    \begin{center}
    \includegraphics[width=\columnwidth]{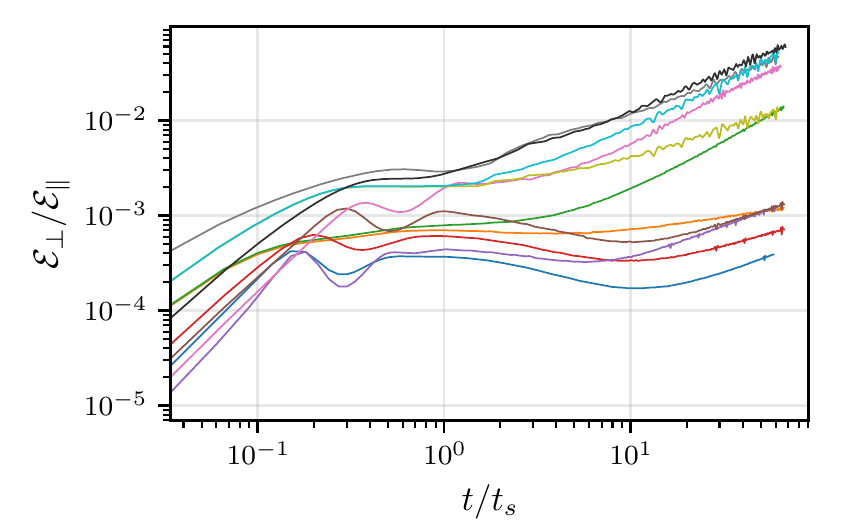}
    \end{center}
    \caption{\label{fig:lontranenA}
    Figure~\ref{fig:lontranen} but all runs
    featured in table~\ref{tab:table3} are included. The high
    Reynolds number runs in the range 160-230, and the low Reynolds number runs
    in the range 10-60 are clearly separated into two groups by at least an
    order of magnitude, with the high Re runs having higher transverse kinetic
    energy. The exception is the low Re bulk viscosity only run (ID 3, green),
    which joins the high Re curves in the end. It has identical initial
    conditions to run 2 (orange) apart from the viscosity type.}
    \end{figure}

\clearpage
\bibliography{2D_Fluid.bib}

\providecommand{\noopsort}[1]{}\providecommand{\singleletter}[1]{#1}%
\begin{thebibliography}{53}%
\makeatletter
\providecommand \@ifxundefined [1]{%
 \@ifx{#1\undefined}
}%
\providecommand \@ifnum [1]{%
 \ifnum #1\expandafter \@firstoftwo
 \else \expandafter \@secondoftwo
 \fi
}%
\providecommand \@ifx [1]{%
 \ifx #1\expandafter \@firstoftwo
 \else \expandafter \@secondoftwo
 \fi
}%
\providecommand \natexlab [1]{#1}%
\providecommand \enquote  [1]{``#1''}%
\providecommand \bibnamefont  [1]{#1}%
\providecommand \bibfnamefont [1]{#1}%
\providecommand \citenamefont [1]{#1}%
\providecommand \href@noop [0]{\@secondoftwo}%
\providecommand \href [0]{\begingroup \@sanitize@url \@href}%
\providecommand \@href[1]{\@@startlink{#1}\@@href}%
\providecommand \@@href[1]{\endgroup#1\@@endlink}%
\providecommand \@sanitize@url [0]{\catcode `\\12\catcode `\$12\catcode
  `\&12\catcode `\#12\catcode `\^12\catcode `\_12\catcode `\%12\relax}%
\providecommand \@@startlink[1]{}%
\providecommand \@@endlink[0]{}%
\providecommand \url  [0]{\begingroup\@sanitize@url \@url }%
\providecommand \@url [1]{\endgroup\@href {#1}{\urlprefix }}%
\providecommand \urlprefix  [0]{URL }%
\providecommand \Eprint [0]{\href }%
\providecommand \doibase [0]{https://doi.org/}%
\providecommand \selectlanguage [0]{\@gobble}%
\providecommand \bibinfo  [0]{\@secondoftwo}%
\providecommand \bibfield  [0]{\@secondoftwo}%
\providecommand \translation [1]{[#1]}%
\providecommand \BibitemOpen [0]{}%
\providecommand \bibitemStop [0]{}%
\providecommand \bibitemNoStop [0]{.\EOS\space}%
\providecommand \EOS [0]{\spacefactor3000\relax}%
\providecommand \BibitemShut  [1]{\csname bibitem#1\endcsname}%
\let\auto@bib@innerbib\@empty
\bibitem [{\citenamefont {Abbott}\ \emph {et~al.}(2016)\citenamefont {Abbott}
  \emph {et~al.}}]{LIGOScientific:2016aoc}%
  \BibitemOpen
  \bibfield  {author} {\bibinfo {author} {\bibfnamefont {B.~P.}\ \bibnamefont
  {Abbott}} \emph {et~al.} (\bibinfo {collaboration} {LIGO Scientific,
  Virgo}),\ }\bibfield  {title} {\bibinfo {title} {{Observation of
  Gravitational Waves from a Binary Black Hole Merger}},\ }\href
  {https://doi.org/10.1103/PhysRevLett.116.061102} {\bibfield  {journal}
  {\bibinfo  {journal} {Phys. Rev. Lett.}\ }\textbf {\bibinfo {volume} {116}},\
  \bibinfo {pages} {061102} (\bibinfo {year} {2016})},\ \Eprint
  {https://arxiv.org/abs/1602.03837} {arXiv:1602.03837 [gr-qc]} \BibitemShut
  {NoStop}%
\bibitem [{\citenamefont {Ricciardone}(2017)}]{Ricciardone:2016ddg}%
  \BibitemOpen
  \bibfield  {author} {\bibinfo {author} {\bibfnamefont {A.}~\bibnamefont
  {Ricciardone}},\ }\bibfield  {title} {\bibinfo {title} {{Primordial
  Gravitational Waves with LISA}},\ }\href
  {https://doi.org/10.1088/1742-6596/840/1/012030} {\bibfield  {journal}
  {\bibinfo  {journal} {J. Phys. Conf. Ser.}\ }\textbf {\bibinfo {volume}
  {840}},\ \bibinfo {pages} {012030} (\bibinfo {year} {2017})},\ \Eprint
  {https://arxiv.org/abs/1612.06799} {arXiv:1612.06799 [astro-ph.CO]}
  \BibitemShut {NoStop}%
\bibitem [{\citenamefont {Christensen}(2019)}]{Christensen:2018iqi}%
  \BibitemOpen
  \bibfield  {author} {\bibinfo {author} {\bibfnamefont {N.}~\bibnamefont
  {Christensen}},\ }\bibfield  {title} {\bibinfo {title} {{Stochastic
  Gravitational Wave Backgrounds}},\ }\href
  {https://doi.org/10.1088/1361-6633/aae6b5} {\bibfield  {journal} {\bibinfo
  {journal} {Rept. Prog. Phys.}\ }\textbf {\bibinfo {volume} {82}},\ \bibinfo
  {pages} {016903} (\bibinfo {year} {2019})},\ \Eprint
  {https://arxiv.org/abs/1811.08797} {arXiv:1811.08797 [gr-qc]} \BibitemShut
  {NoStop}%
\bibitem [{\citenamefont {Caprini}\ and\ \citenamefont
  {Figueroa}(2018)}]{Caprini:2018mtu}%
  \BibitemOpen
  \bibfield  {author} {\bibinfo {author} {\bibfnamefont {C.}~\bibnamefont
  {Caprini}}\ and\ \bibinfo {author} {\bibfnamefont {D.~G.}\ \bibnamefont
  {Figueroa}},\ }\bibfield  {title} {\bibinfo {title} {{Cosmological
  Backgrounds of Gravitational Waves}},\ }\href
  {https://doi.org/10.1088/1361-6382/aac608} {\bibfield  {journal} {\bibinfo
  {journal} {Class. Quant. Grav.}\ }\textbf {\bibinfo {volume} {35}},\ \bibinfo
  {pages} {163001} (\bibinfo {year} {2018})},\ \Eprint
  {https://arxiv.org/abs/1801.04268} {arXiv:1801.04268 [astro-ph.CO]}
  \BibitemShut {NoStop}%
\bibitem [{\citenamefont {Allen}(1996)}]{Allen:1996vm}%
  \BibitemOpen
  \bibfield  {author} {\bibinfo {author} {\bibfnamefont {B.}~\bibnamefont
  {Allen}},\ }\bibfield  {title} {\bibinfo {title} {{The Stochastic gravity
  wave background: Sources and detection}},\ }in\ \href@noop {} {\emph
  {\bibinfo {booktitle} {{Les Houches School of Physics: Astrophysical Sources
  of Gravitational Radiation}}}}\ (\bibinfo {year} {1996})\ pp.\ \bibinfo
  {pages} {373--417},\ \Eprint {https://arxiv.org/abs/gr-qc/9604033}
  {arXiv:gr-qc/9604033} \BibitemShut {NoStop}%
\bibitem [{\citenamefont {Maggiore}(2000)}]{Maggiore:1999vm}%
  \BibitemOpen
  \bibfield  {author} {\bibinfo {author} {\bibfnamefont {M.}~\bibnamefont
  {Maggiore}},\ }\bibfield  {title} {\bibinfo {title} {{Gravitational wave
  experiments and early universe cosmology}},\ }\href
  {https://doi.org/10.1016/S0370-1573(99)00102-7} {\bibfield  {journal}
  {\bibinfo  {journal} {Phys. Rept.}\ }\textbf {\bibinfo {volume} {331}},\
  \bibinfo {pages} {283} (\bibinfo {year} {2000})},\ \Eprint
  {https://arxiv.org/abs/gr-qc/9909001} {arXiv:gr-qc/9909001} \BibitemShut
  {NoStop}%
\bibitem [{\citenamefont {Amaro-Seoane}\ \emph {et~al.}(2017)\citenamefont
  {Amaro-Seoane} \emph {et~al.}}]{LISA:2017pwj}%
  \BibitemOpen
  \bibfield  {author} {\bibinfo {author} {\bibfnamefont {P.}~\bibnamefont
  {Amaro-Seoane}} \emph {et~al.} (\bibinfo {collaboration} {LISA}),\ }\bibfield
   {title} {\bibinfo {title} {{Laser Interferometer Space Antenna}},\
  }\href@noop {} {\  (\bibinfo {year} {2017})},\ \Eprint
  {https://arxiv.org/abs/1702.00786} {arXiv:1702.00786 [astro-ph.IM]}
  \BibitemShut {NoStop}%
\bibitem [{\citenamefont {Witten}(1984)}]{Witten:1984rs}%
  \BibitemOpen
  \bibfield  {author} {\bibinfo {author} {\bibfnamefont {E.}~\bibnamefont
  {Witten}},\ }\bibfield  {title} {\bibinfo {title} {{Cosmic Separation of
  Phases}},\ }\href {https://doi.org/10.1103/PhysRevD.30.272} {\bibfield
  {journal} {\bibinfo  {journal} {Phys.Rev.}\ }\textbf {\bibinfo {volume}
  {D30}},\ \bibinfo {pages} {272} (\bibinfo {year} {1984})}\BibitemShut
  {NoStop}%
\bibitem [{\citenamefont {Hogan}(1986)}]{Hogan:1986qda}%
  \BibitemOpen
  \bibfield  {author} {\bibinfo {author} {\bibfnamefont {C.~J.}\ \bibnamefont
  {Hogan}},\ }\bibfield  {title} {\bibinfo {title} {{Gravitational radiation
  from cosmological phase transitions}},\ }\href@noop {} {\bibfield  {journal}
  {\bibinfo  {journal} {Mon. Not. Roy. Astron. Soc.}\ }\textbf {\bibinfo
  {volume} {218}},\ \bibinfo {pages} {629} (\bibinfo {year}
  {1986})}\BibitemShut {NoStop}%
\bibitem [{\citenamefont {Kamionkowski}\ \emph {et~al.}(1994)\citenamefont
  {Kamionkowski}, \citenamefont {Kosowsky},\ and\ \citenamefont
  {Turner}}]{Kamionkowski:1993fg}%
  \BibitemOpen
  \bibfield  {author} {\bibinfo {author} {\bibfnamefont {M.}~\bibnamefont
  {Kamionkowski}}, \bibinfo {author} {\bibfnamefont {A.}~\bibnamefont
  {Kosowsky}},\ and\ \bibinfo {author} {\bibfnamefont {M.~S.}\ \bibnamefont
  {Turner}},\ }\bibfield  {title} {\bibinfo {title} {{Gravitational radiation
  from first order phase transitions}},\ }\href
  {https://doi.org/10.1103/PhysRevD.49.2837} {\bibfield  {journal} {\bibinfo
  {journal} {Phys. Rev. D}\ }\textbf {\bibinfo {volume} {49}},\ \bibinfo
  {pages} {2837} (\bibinfo {year} {1994})},\ \Eprint
  {https://arxiv.org/abs/astro-ph/9310044} {arXiv:astro-ph/9310044}
  \BibitemShut {NoStop}%
\bibitem [{\citenamefont {Caprini}\ \emph {et~al.}(2020)\citenamefont {Caprini}
  \emph {et~al.}}]{Caprini:2019egz}%
  \BibitemOpen
  \bibfield  {author} {\bibinfo {author} {\bibfnamefont {C.}~\bibnamefont
  {Caprini}} \emph {et~al.},\ }\bibfield  {title} {\bibinfo {title} {{Detecting
  gravitational waves from cosmological phase transitions with LISA: an
  update}},\ }\href {https://doi.org/10.1088/1475-7516/2020/03/024} {\bibfield
  {journal} {\bibinfo  {journal} {JCAP}\ }\textbf {\bibinfo {volume} {03}},\
  \bibinfo {pages} {024}},\ \Eprint {https://arxiv.org/abs/1910.13125}
  {arXiv:1910.13125 [astro-ph.CO]} \BibitemShut {NoStop}%
\bibitem [{\citenamefont {Guth}\ and\ \citenamefont
  {Weinberg}(1981)}]{Guth:1981uk}%
  \BibitemOpen
  \bibfield  {author} {\bibinfo {author} {\bibfnamefont {A.~H.}\ \bibnamefont
  {Guth}}\ and\ \bibinfo {author} {\bibfnamefont {E.~J.}\ \bibnamefont
  {Weinberg}},\ }\bibfield  {title} {\bibinfo {title} {{Cosmological
  Consequences of a First Order Phase Transition in the SU(5) Grand Unified
  Model}},\ }\href {https://doi.org/10.1103/PhysRevD.23.876} {\bibfield
  {journal} {\bibinfo  {journal} {Phys. Rev.}\ }\textbf {\bibinfo {volume}
  {D23}},\ \bibinfo {pages} {876} (\bibinfo {year} {1981})}\BibitemShut
  {NoStop}%
\bibitem [{\citenamefont {Steinhardt}(1982)}]{Steinhardt:1981ct}%
  \BibitemOpen
  \bibfield  {author} {\bibinfo {author} {\bibfnamefont {P.~J.}\ \bibnamefont
  {Steinhardt}},\ }\bibfield  {title} {\bibinfo {title} {{Relativistic
  detonation waves and bubble growth in false vacuum decay}},\ }\href
  {https://doi.org/10.1103/PhysRevD.25.2074} {\bibfield  {journal} {\bibinfo
  {journal} {Phys.Rev.}\ }\textbf {\bibinfo {volume} {D25}},\ \bibinfo {pages}
  {2074} (\bibinfo {year} {1982})}\BibitemShut {NoStop}%
\bibitem [{\citenamefont {Ignatius}\ \emph {et~al.}(1994)\citenamefont
  {Ignatius}, \citenamefont {Kajantie}, \citenamefont {Kurki-Suonio},\ and\
  \citenamefont {Laine}}]{Ignatius:1993qn}%
  \BibitemOpen
  \bibfield  {author} {\bibinfo {author} {\bibfnamefont {J.}~\bibnamefont
  {Ignatius}}, \bibinfo {author} {\bibfnamefont {K.}~\bibnamefont {Kajantie}},
  \bibinfo {author} {\bibfnamefont {H.}~\bibnamefont {Kurki-Suonio}},\ and\
  \bibinfo {author} {\bibfnamefont {M.}~\bibnamefont {Laine}},\ }\bibfield
  {title} {\bibinfo {title} {{The growth of bubbles in cosmological phase
  transitions}},\ }\href {https://doi.org/10.1103/PhysRevD.49.3854} {\bibfield
  {journal} {\bibinfo  {journal} {Phys. Rev. D}\ }\textbf {\bibinfo {volume}
  {49}},\ \bibinfo {pages} {3854} (\bibinfo {year} {1994})},\ \Eprint
  {https://arxiv.org/abs/astro-ph/9309059} {arXiv:astro-ph/9309059}
  \BibitemShut {NoStop}%
\bibitem [{\citenamefont {Espinosa}\ \emph {et~al.}(2010)\citenamefont
  {Espinosa}, \citenamefont {Konstandin}, \citenamefont {No},\ and\
  \citenamefont {Servant}}]{Espinosa:2010hh}%
  \BibitemOpen
  \bibfield  {author} {\bibinfo {author} {\bibfnamefont {J.~R.}\ \bibnamefont
  {Espinosa}}, \bibinfo {author} {\bibfnamefont {T.}~\bibnamefont
  {Konstandin}}, \bibinfo {author} {\bibfnamefont {J.~M.}\ \bibnamefont {No}},\
  and\ \bibinfo {author} {\bibfnamefont {G.}~\bibnamefont {Servant}},\
  }\bibfield  {title} {\bibinfo {title} {{Energy Budget of Cosmological
  First-order Phase Transitions}},\ }\href
  {https://doi.org/10.1088/1475-7516/2010/06/028} {\bibfield  {journal}
  {\bibinfo  {journal} {JCAP}\ }\textbf {\bibinfo {volume} {06}},\ \bibinfo
  {pages} {028}},\ \Eprint {https://arxiv.org/abs/1004.4187} {arXiv:1004.4187
  [hep-ph]} \BibitemShut {NoStop}%
\bibitem [{\citenamefont {M\'egevand}\ and\ \citenamefont
  {Ram\'\i{}rez}(2018)}]{Megevand:2017vtb}%
  \BibitemOpen
  \bibfield  {author} {\bibinfo {author} {\bibfnamefont {A.}~\bibnamefont
  {M\'egevand}}\ and\ \bibinfo {author} {\bibfnamefont {S.}~\bibnamefont
  {Ram\'\i{}rez}},\ }\bibfield  {title} {\bibinfo {title} {{Bubble nucleation
  and growth in slow cosmological phase transitions}},\ }\href
  {https://doi.org/10.1016/j.nuclphysb.2018.01.012} {\bibfield  {journal}
  {\bibinfo  {journal} {Nucl. Phys. B}\ }\textbf {\bibinfo {volume} {928}},\
  \bibinfo {pages} {38} (\bibinfo {year} {2018})},\ \Eprint
  {https://arxiv.org/abs/1710.06279} {arXiv:1710.06279 [astro-ph.CO]}
  \BibitemShut {NoStop}%
\bibitem [{\citenamefont {Hindmarsh}\ \emph {et~al.}(2014)\citenamefont
  {Hindmarsh}, \citenamefont {Huber}, \citenamefont {Rummukainen},\ and\
  \citenamefont {Weir}}]{Hindmarsh:2013xza}%
  \BibitemOpen
  \bibfield  {author} {\bibinfo {author} {\bibfnamefont {M.}~\bibnamefont
  {Hindmarsh}}, \bibinfo {author} {\bibfnamefont {S.~J.}\ \bibnamefont
  {Huber}}, \bibinfo {author} {\bibfnamefont {K.}~\bibnamefont {Rummukainen}},\
  and\ \bibinfo {author} {\bibfnamefont {D.~J.}\ \bibnamefont {Weir}},\
  }\bibfield  {title} {\bibinfo {title} {{Gravitational waves from the sound of
  a first order phase transition}},\ }\href
  {https://doi.org/10.1103/PhysRevLett.112.041301} {\bibfield  {journal}
  {\bibinfo  {journal} {Phys. Rev. Lett.}\ }\textbf {\bibinfo {volume} {112}},\
  \bibinfo {pages} {041301} (\bibinfo {year} {2014})},\ \Eprint
  {https://arxiv.org/abs/1304.2433} {arXiv:1304.2433 [hep-ph]} \BibitemShut
  {NoStop}%
\bibitem [{\citenamefont {Hindmarsh}\ \emph {et~al.}(2015)\citenamefont
  {Hindmarsh}, \citenamefont {Huber}, \citenamefont {Rummukainen},\ and\
  \citenamefont {Weir}}]{Hindmarsh:2015qta}%
  \BibitemOpen
  \bibfield  {author} {\bibinfo {author} {\bibfnamefont {M.}~\bibnamefont
  {Hindmarsh}}, \bibinfo {author} {\bibfnamefont {S.~J.}\ \bibnamefont
  {Huber}}, \bibinfo {author} {\bibfnamefont {K.}~\bibnamefont {Rummukainen}},\
  and\ \bibinfo {author} {\bibfnamefont {D.~J.}\ \bibnamefont {Weir}},\
  }\bibfield  {title} {\bibinfo {title} {{Numerical simulations of acoustically
  generated gravitational waves at a first order phase transition}},\ }\href
  {https://doi.org/10.1103/PhysRevD.92.123009} {\bibfield  {journal} {\bibinfo
  {journal} {Phys. Rev. D}\ }\textbf {\bibinfo {volume} {92}},\ \bibinfo
  {pages} {123009} (\bibinfo {year} {2015})},\ \Eprint
  {https://arxiv.org/abs/1504.03291} {arXiv:1504.03291 [astro-ph.CO]}
  \BibitemShut {NoStop}%
\bibitem [{\citenamefont {Hindmarsh}\ \emph {et~al.}(2017)\citenamefont
  {Hindmarsh}, \citenamefont {Huber}, \citenamefont {Rummukainen},\ and\
  \citenamefont {Weir}}]{Hindmarsh:2017gnf}%
  \BibitemOpen
  \bibfield  {author} {\bibinfo {author} {\bibfnamefont {M.}~\bibnamefont
  {Hindmarsh}}, \bibinfo {author} {\bibfnamefont {S.~J.}\ \bibnamefont
  {Huber}}, \bibinfo {author} {\bibfnamefont {K.}~\bibnamefont {Rummukainen}},\
  and\ \bibinfo {author} {\bibfnamefont {D.~J.}\ \bibnamefont {Weir}},\
  }\bibfield  {title} {\bibinfo {title} {{Shape of the acoustic gravitational
  wave power spectrum from a first order phase transition}},\ }\href
  {https://doi.org/10.1103/PhysRevD.96.103520} {\bibfield  {journal} {\bibinfo
  {journal} {Phys. Rev. D}\ }\textbf {\bibinfo {volume} {96}},\ \bibinfo
  {pages} {103520} (\bibinfo {year} {2017})},\ \bibinfo {note} {[Erratum:
  Phys.Rev.D 101, 089902(E) (2020)]},\ \Eprint
  {https://arxiv.org/abs/1704.05871} {arXiv:1704.05871 [astro-ph.CO]}
  \BibitemShut {NoStop}%
\bibitem [{\citenamefont {Hindmarsh}\ and\ \citenamefont
  {Hijazi}(2019)}]{Hindmarsh:2019phv}%
  \BibitemOpen
  \bibfield  {author} {\bibinfo {author} {\bibfnamefont {M.}~\bibnamefont
  {Hindmarsh}}\ and\ \bibinfo {author} {\bibfnamefont {M.}~\bibnamefont
  {Hijazi}},\ }\bibfield  {title} {\bibinfo {title} {{Gravitational waves from
  first order cosmological phase transitions in the Sound Shell Model}},\
  }\href {https://doi.org/10.1088/1475-7516/2019/12/062} {\bibfield  {journal}
  {\bibinfo  {journal} {JCAP}\ }\textbf {\bibinfo {volume} {12}},\ \bibinfo
  {pages} {062}},\ \Eprint {https://arxiv.org/abs/1909.10040} {arXiv:1909.10040
  [astro-ph.CO]} \BibitemShut {NoStop}%
\bibitem [{\citenamefont {Cutting}\ \emph {et~al.}(2019)\citenamefont
  {Cutting}, \citenamefont {Hindmarsh},\ and\ \citenamefont
  {Weir}}]{Cutting:2019zws}%
  \BibitemOpen
  \bibfield  {author} {\bibinfo {author} {\bibfnamefont {D.}~\bibnamefont
  {Cutting}}, \bibinfo {author} {\bibfnamefont {M.}~\bibnamefont {Hindmarsh}},\
  and\ \bibinfo {author} {\bibfnamefont {D.~J.}\ \bibnamefont {Weir}},\
  }\bibfield  {title} {\bibinfo {title} {{Vorticity, kinetic energy, and
  suppressed gravitational wave production in strong first order phase
  transitions}},\ }\href@noop {} {\  (\bibinfo {year} {2019})},\ \Eprint
  {https://arxiv.org/abs/1906.00480} {arXiv:1906.00480 [hep-ph]} \BibitemShut
  {NoStop}%
\bibitem [{\citenamefont {L'vov}\ \emph {et~al.}(1997)\citenamefont {L'vov},
  \citenamefont {L'vov}, \citenamefont {Newell},\ and\ \citenamefont
  {Zakharov}}]{L_vov_1997}%
  \BibitemOpen
  \bibfield  {author} {\bibinfo {author} {\bibfnamefont {V.~S.}\ \bibnamefont
  {L'vov}}, \bibinfo {author} {\bibfnamefont {Y.}~\bibnamefont {L'vov}},
  \bibinfo {author} {\bibfnamefont {A.~C.}\ \bibnamefont {Newell}},\ and\
  \bibinfo {author} {\bibfnamefont {V.}~\bibnamefont {Zakharov}},\ }\bibfield
  {title} {\bibinfo {title} {Statistical description of acoustic turbulence},\
  }\href {https://doi.org/10.1103/physreve.56.390} {\bibfield  {journal}
  {\bibinfo  {journal} {Physical Review E}\ }\textbf {\bibinfo {volume} {56}},\
  \bibinfo {pages} {390} (\bibinfo {year} {1997})}\BibitemShut {NoStop}%
\bibitem [{\citenamefont {L'vov}\ \emph {et~al.}(2000)\citenamefont {L'vov},
  \citenamefont {L'vov},\ and\ \citenamefont {Pomyalov}}]{Lvov:2000bdb}%
  \BibitemOpen
  \bibfield  {author} {\bibinfo {author} {\bibfnamefont {V.~S.}\ \bibnamefont
  {L'vov}}, \bibinfo {author} {\bibfnamefont {Y.~V.}\ \bibnamefont {L'vov}},\
  and\ \bibinfo {author} {\bibfnamefont {A.}~\bibnamefont {Pomyalov}},\
  }\bibfield  {title} {\bibinfo {title} {{Anisotropic spectra of acoustic
  turbulence}},\ }\href {https://doi.org/10.1103/PhysRevE.61.2586} {\bibfield
  {journal} {\bibinfo  {journal} {Phys. Rev. E}\ }\textbf {\bibinfo {volume}
  {61}},\ \bibinfo {pages} {2586} (\bibinfo {year} {2000})},\ \Eprint
  {https://arxiv.org/abs/chao-dyn/9905032} {arXiv:chao-dyn/9905032}
  \BibitemShut {NoStop}%
\bibitem [{\citenamefont {Frisch}\ and\ \citenamefont
  {Bec}(2000)}]{Frisch:2000td}%
  \BibitemOpen
  \bibfield  {author} {\bibinfo {author} {\bibfnamefont {U.}~\bibnamefont
  {Frisch}}\ and\ \bibinfo {author} {\bibfnamefont {J.}~\bibnamefont {Bec}},\
  }\bibfield  {title} {\bibinfo {title} {{'Burgulence'}},\ }in\ \href@noop {}
  {\emph {\bibinfo {booktitle} {{Les Houches 2000 Summer School: Session 74:
  New Trends in Turbulence}}}}\ (\bibinfo {year} {2000})\ \Eprint
  {https://arxiv.org/abs/nlin/0012033} {arXiv:nlin/0012033} \BibitemShut
  {NoStop}%
\bibitem [{\citenamefont {Bec}\ and\ \citenamefont {Khanin}(2007)}]{BEC_2007}%
  \BibitemOpen
  \bibfield  {author} {\bibinfo {author} {\bibfnamefont {J.}~\bibnamefont
  {Bec}}\ and\ \bibinfo {author} {\bibfnamefont {K.}~\bibnamefont {Khanin}},\
  }\bibfield  {title} {\bibinfo {title} {Burgers turbulence},\ }\href
  {https://doi.org/10.1016/j.physrep.2007.04.002} {\bibfield  {journal}
  {\bibinfo  {journal} {Physics Reports}\ }\textbf {\bibinfo {volume} {447}},\
  \bibinfo {pages} {1} (\bibinfo {year} {2007})}\BibitemShut {NoStop}%
\bibitem [{\citenamefont {Porter}\ \emph {et~al.}(1992)\citenamefont {Porter},
  \citenamefont {Pouquet},\ and\ \citenamefont {Woodward}}]{Porter1992ANS}%
  \BibitemOpen
  \bibfield  {author} {\bibinfo {author} {\bibfnamefont {D.}~\bibnamefont
  {Porter}}, \bibinfo {author} {\bibfnamefont {A.}~\bibnamefont {Pouquet}},\
  and\ \bibinfo {author} {\bibfnamefont {P.}~\bibnamefont {Woodward}},\
  }\bibfield  {title} {\bibinfo {title} {A numerical study of supersonic
  turbulence},\ }\href@noop {} {\bibfield  {journal} {\bibinfo  {journal}
  {Theoretical and Computational Fluid Dynamics}\ }\textbf {\bibinfo {volume}
  {4}},\ \bibinfo {pages} {13} (\bibinfo {year} {1992})}\BibitemShut {NoStop}%
\bibitem [{\citenamefont {Kida}\ and\ \citenamefont
  {Orszag}(1992)}]{Kida1992EnergyAS}%
  \BibitemOpen
  \bibfield  {author} {\bibinfo {author} {\bibfnamefont {S.}~\bibnamefont
  {Kida}}\ and\ \bibinfo {author} {\bibfnamefont {S.}~\bibnamefont {Orszag}},\
  }\bibfield  {title} {\bibinfo {title} {Energy and spectral dynamics in
  decaying compressible turbulence},\ }\href@noop {} {\bibfield  {journal}
  {\bibinfo  {journal} {Journal of Scientific Computing}\ }\textbf {\bibinfo
  {volume} {7}},\ \bibinfo {pages} {1} (\bibinfo {year} {1992})}\BibitemShut
  {NoStop}%
\bibitem [{\citenamefont {Ducros}\ \emph {et~al.}(1999)\citenamefont {Ducros},
  \citenamefont {Ferrand}, \citenamefont {Nicoud}, \citenamefont {Weber},
  \citenamefont {Darracq}, \citenamefont {Gacherieu},\ and\ \citenamefont
  {Poinsot}}]{DUCROS1999517}%
  \BibitemOpen
  \bibfield  {author} {\bibinfo {author} {\bibfnamefont {F.}~\bibnamefont
  {Ducros}}, \bibinfo {author} {\bibfnamefont {V.}~\bibnamefont {Ferrand}},
  \bibinfo {author} {\bibfnamefont {F.}~\bibnamefont {Nicoud}}, \bibinfo
  {author} {\bibfnamefont {C.}~\bibnamefont {Weber}}, \bibinfo {author}
  {\bibfnamefont {D.}~\bibnamefont {Darracq}}, \bibinfo {author} {\bibfnamefont
  {C.}~\bibnamefont {Gacherieu}},\ and\ \bibinfo {author} {\bibfnamefont
  {T.}~\bibnamefont {Poinsot}},\ }\bibfield  {title} {\bibinfo {title}
  {Large-eddy simulation of the shock/turbulence interaction},\ }\href
  {https://doi.org/https://doi.org/10.1006/jcph.1999.6238} {\bibfield
  {journal} {\bibinfo  {journal} {Journal of Computational Physics}\ }\textbf
  {\bibinfo {volume} {152}},\ \bibinfo {pages} {517} (\bibinfo {year}
  {1999})}\BibitemShut {NoStop}%
\bibitem [{\citenamefont {Pen}\ and\ \citenamefont
  {Turok}(2016)}]{Pen:2015qta}%
  \BibitemOpen
  \bibfield  {author} {\bibinfo {author} {\bibfnamefont {U.-L.}\ \bibnamefont
  {Pen}}\ and\ \bibinfo {author} {\bibfnamefont {N.}~\bibnamefont {Turok}},\
  }\bibfield  {title} {\bibinfo {title} {{Shocks in the Early Universe}},\
  }\href {https://doi.org/10.1103/PhysRevLett.117.131301} {\bibfield  {journal}
  {\bibinfo  {journal} {Phys. Rev. Lett.}\ }\textbf {\bibinfo {volume} {117}},\
  \bibinfo {pages} {131301} (\bibinfo {year} {2016})},\ \Eprint
  {https://arxiv.org/abs/1510.02985} {arXiv:1510.02985 [astro-ph.CO]}
  \BibitemShut {NoStop}%
\bibitem [{\citenamefont {Roper~Pol}\ \emph {et~al.}(2020)\citenamefont
  {Roper~Pol}, \citenamefont {Mandal}, \citenamefont {Brandenburg},
  \citenamefont {Kahniashvili},\ and\ \citenamefont
  {Kosowsky}}]{RoperPol:2019wvy}%
  \BibitemOpen
  \bibfield  {author} {\bibinfo {author} {\bibfnamefont {A.}~\bibnamefont
  {Roper~Pol}}, \bibinfo {author} {\bibfnamefont {S.}~\bibnamefont {Mandal}},
  \bibinfo {author} {\bibfnamefont {A.}~\bibnamefont {Brandenburg}}, \bibinfo
  {author} {\bibfnamefont {T.}~\bibnamefont {Kahniashvili}},\ and\ \bibinfo
  {author} {\bibfnamefont {A.}~\bibnamefont {Kosowsky}},\ }\bibfield  {title}
  {\bibinfo {title} {{Numerical simulations of gravitational waves from
  early-universe turbulence}},\ }\href
  {https://doi.org/10.1103/PhysRevD.102.083512} {\bibfield  {journal} {\bibinfo
   {journal} {Phys. Rev. D}\ }\textbf {\bibinfo {volume} {102}},\ \bibinfo
  {pages} {083512} (\bibinfo {year} {2020})},\ \Eprint
  {https://arxiv.org/abs/1903.08585} {arXiv:1903.08585 [astro-ph.CO]}
  \BibitemShut {NoStop}%
\bibitem [{\citenamefont {Brandenburg}\ \emph {et~al.}(1996)\citenamefont
  {Brandenburg}, \citenamefont {Enqvist},\ and\ \citenamefont
  {Olesen}}]{Brandenburg:1996fc}%
  \BibitemOpen
  \bibfield  {author} {\bibinfo {author} {\bibfnamefont {A.}~\bibnamefont
  {Brandenburg}}, \bibinfo {author} {\bibfnamefont {K.}~\bibnamefont
  {Enqvist}},\ and\ \bibinfo {author} {\bibfnamefont {P.}~\bibnamefont
  {Olesen}},\ }\bibfield  {title} {\bibinfo {title} {{Large scale magnetic
  fields from hydromagnetic turbulence in the very early universe}},\ }\href
  {https://doi.org/10.1103/PhysRevD.54.1291} {\bibfield  {journal} {\bibinfo
  {journal} {Phys. Rev. D}\ }\textbf {\bibinfo {volume} {54}},\ \bibinfo
  {pages} {1291} (\bibinfo {year} {1996})},\ \Eprint
  {https://arxiv.org/abs/astro-ph/9602031} {arXiv:astro-ph/9602031}
  \BibitemShut {NoStop}%
\bibitem [{\citenamefont {Arnold}\ \emph {et~al.}(2006)\citenamefont {Arnold},
  \citenamefont {Dogan},\ and\ \citenamefont {Moore}}]{Arnold:2006fz}%
  \BibitemOpen
  \bibfield  {author} {\bibinfo {author} {\bibfnamefont {P.~B.}\ \bibnamefont
  {Arnold}}, \bibinfo {author} {\bibfnamefont {C.}~\bibnamefont {Dogan}},\ and\
  \bibinfo {author} {\bibfnamefont {G.~D.}\ \bibnamefont {Moore}},\ }\bibfield
  {title} {\bibinfo {title} {{The Bulk Viscosity of High-Temperature QCD}},\
  }\href {https://doi.org/10.1103/PhysRevD.74.085021} {\bibfield  {journal}
  {\bibinfo  {journal} {Phys. Rev. D}\ }\textbf {\bibinfo {volume} {74}},\
  \bibinfo {pages} {085021} (\bibinfo {year} {2006})},\ \Eprint
  {https://arxiv.org/abs/hep-ph/0608012} {arXiv:hep-ph/0608012} \BibitemShut
  {NoStop}%
\bibitem [{\citenamefont {Hewitt}\ and\ \citenamefont {Hewitt}(1979)}]{Gibbs}%
  \BibitemOpen
  \bibfield  {author} {\bibinfo {author} {\bibfnamefont {E.}~\bibnamefont
  {Hewitt}}\ and\ \bibinfo {author} {\bibfnamefont {R.~E.}\ \bibnamefont
  {Hewitt}},\ }\bibfield  {title} {\bibinfo {title} {The gibbs-wilbraham
  phenomenon: An episode in fourier analysis},\ }\href
  {http://www.jstor.org/stable/41133555} {\bibfield  {journal} {\bibinfo
  {journal} {Archive for History of Exact Sciences}\ }\textbf {\bibinfo
  {volume} {21}},\ \bibinfo {pages} {129} (\bibinfo {year} {1979})}\BibitemShut
  {NoStop}%
\bibitem [{\citenamefont {Burgers}(1948)}]{Burgers_1948}%
  \BibitemOpen
  \bibfield  {author} {\bibinfo {author} {\bibfnamefont {J.}~\bibnamefont
  {Burgers}},\ }\bibfield  {title} {\bibinfo {title} {A mathematical model
  illustrating the theory of turbulence},\ }\href
  {https://doi.org/10.1016/s0065-2156(08)70100-5} {\bibfield  {journal}
  {\bibinfo  {journal} {Advances in Applied Mechanics}\ }\textbf {\bibinfo
  {volume} {1}},\ \bibinfo {pages} {171} (\bibinfo {year} {1948})}\BibitemShut
  {NoStop}%
\bibitem [{\citenamefont {Kadomtsev}\ and\ \citenamefont
  {Petviashvilil}(1973)}]{KP}%
  \BibitemOpen
  \bibfield  {author} {\bibinfo {author} {\bibfnamefont {B.}~\bibnamefont
  {Kadomtsev}}\ and\ \bibinfo {author} {\bibfnamefont {V.}~\bibnamefont
  {Petviashvilil}},\ }\href@noop {} {\bibfield  {journal} {\bibinfo  {journal}
  {Dokl. Akad. Nauk SSSR}\ }\textbf {\bibinfo {volume} {208}},\ \bibinfo
  {pages} {794} (\bibinfo {year} {1973})}\BibitemShut {NoStop}%
\bibitem [{\citenamefont {Kuznetsov}\ and\ \citenamefont
  {Krasnoselskikh}(2008)}]{Kuznetsov}%
  \BibitemOpen
  \bibfield  {author} {\bibinfo {author} {\bibfnamefont {E.}~\bibnamefont
  {Kuznetsov}}\ and\ \bibinfo {author} {\bibfnamefont {V.}~\bibnamefont
  {Krasnoselskikh}},\ }\bibfield  {title} {\bibinfo {title} {Anisotropic
  spectra of acoustic type turbulence},\ }\href
  {https://doi.org/10.1063/1.2928160} {\bibfield  {journal} {\bibinfo
  {journal} {Physics of Plasmas}\ }\textbf {\bibinfo {volume} {15}},\ \bibinfo
  {pages} {062305} (\bibinfo {year} {2008})},\ \Eprint
  {https://arxiv.org/abs/https://doi.org/10.1063/1.2928160}
  {https://doi.org/10.1063/1.2928160} \BibitemShut {NoStop}%
\bibitem [{\citenamefont {Erd{\'e}lyi}\ \emph {et~al.}(1954)\citenamefont
  {Erd{\'e}lyi}, \citenamefont {Magnus}, \citenamefont {Oberhetinger},\ and\
  \citenamefont {Tricomi}}]{Erdelyi}%
  \BibitemOpen
  \bibfield  {author} {\bibinfo {author} {\bibfnamefont {A.}~\bibnamefont
  {Erd{\'e}lyi}}, \bibinfo {author} {\bibfnamefont {W.}~\bibnamefont {Magnus}},
  \bibinfo {author} {\bibfnamefont {F.}~\bibnamefont {Oberhetinger}},\ and\
  \bibinfo {author} {\bibfnamefont {F.}~\bibnamefont {Tricomi}},\ }\href@noop
  {} {\emph {\bibinfo {title} {Tables of integral transforms}}},\ Vol.~\bibinfo
  {volume} {2}\ (\bibinfo  {publisher} {McGraw-Hill},\ \bibinfo {year} {1954})\
  Chap.~\bibinfo {chapter} {13}\BibitemShut {NoStop}%
\bibitem [{\citenamefont {Saffman}(1971)}]{saffman1971}%
  \BibitemOpen
  \bibfield  {author} {\bibinfo {author} {\bibfnamefont {P.~G.}\ \bibnamefont
  {Saffman}},\ }\bibfield  {title} {\bibinfo {title} {On the spectrum and decay
  of random two‐dimensional vorticity distributions at large reynolds
  number},\ }\bibfield  {journal} {\bibinfo  {journal} {Studies in Applied
  Mathematics}\ }\textbf {\bibinfo {volume} {50}},\ \href
  {https://doi.org/10.1002/sapm1971504377} {10.1002/sapm1971504377} (\bibinfo
  {year} {1971})\BibitemShut {NoStop}%
\bibitem [{\citenamefont {Lesieur}(2008)}]{lesieur2008turbulence}%
  \BibitemOpen
  \bibfield  {author} {\bibinfo {author} {\bibfnamefont {M.}~\bibnamefont
  {Lesieur}},\ }\href {https://books.google.fi/books?id=xKUDN22Y7OYC} {\emph
  {\bibinfo {title} {Turbulence in Fluids}}},\ Fluid Mechanics and Its
  Applications\ (\bibinfo  {publisher} {Springer Netherlands},\ \bibinfo {year}
  {2008})\BibitemShut {NoStop}%
\bibitem [{\citenamefont {Olesen}(1997)}]{Olesen:1996ts}%
  \BibitemOpen
  \bibfield  {author} {\bibinfo {author} {\bibfnamefont {P.}~\bibnamefont
  {Olesen}},\ }\bibfield  {title} {\bibinfo {title} {{On inverse cascades in
  astrophysics}},\ }\href {https://doi.org/10.1016/S0370-2693(97)00235-9}
  {\bibfield  {journal} {\bibinfo  {journal} {Phys. Lett. B}\ }\textbf
  {\bibinfo {volume} {398}},\ \bibinfo {pages} {321} (\bibinfo {year}
  {1997})},\ \Eprint {https://arxiv.org/abs/astro-ph/9610154}
  {arXiv:astro-ph/9610154} \BibitemShut {NoStop}%
\bibitem [{\citenamefont {Brandenburg}\ and\ \citenamefont
  {Kahniashvili}(2017)}]{Brandenburg:2016odr}%
  \BibitemOpen
  \bibfield  {author} {\bibinfo {author} {\bibfnamefont {A.}~\bibnamefont
  {Brandenburg}}\ and\ \bibinfo {author} {\bibfnamefont {T.}~\bibnamefont
  {Kahniashvili}},\ }\bibfield  {title} {\bibinfo {title} {{Classes of
  hydrodynamic and magnetohydrodynamic turbulent decay}},\ }\href
  {https://doi.org/10.1103/PhysRevLett.118.055102} {\bibfield  {journal}
  {\bibinfo  {journal} {Phys. Rev. Lett.}\ }\textbf {\bibinfo {volume} {118}},\
  \bibinfo {pages} {055102} (\bibinfo {year} {2017})},\ \Eprint
  {https://arxiv.org/abs/1607.01360} {arXiv:1607.01360 [physics.flu-dyn]}
  \BibitemShut {NoStop}%
\bibitem [{\citenamefont {Kosowsky}\ \emph {et~al.}(2002)\citenamefont
  {Kosowsky}, \citenamefont {Mack},\ and\ \citenamefont
  {Kahniashvili}}]{Kosowsky:2001xp}%
  \BibitemOpen
  \bibfield  {author} {\bibinfo {author} {\bibfnamefont {A.}~\bibnamefont
  {Kosowsky}}, \bibinfo {author} {\bibfnamefont {A.}~\bibnamefont {Mack}},\
  and\ \bibinfo {author} {\bibfnamefont {T.}~\bibnamefont {Kahniashvili}},\
  }\bibfield  {title} {\bibinfo {title} {{Gravitational radiation from
  cosmological turbulence}},\ }\href
  {https://doi.org/10.1103/PhysRevD.66.024030} {\bibfield  {journal} {\bibinfo
  {journal} {Phys. Rev. D}\ }\textbf {\bibinfo {volume} {66}},\ \bibinfo
  {pages} {024030} (\bibinfo {year} {2002})},\ \Eprint
  {https://arxiv.org/abs/astro-ph/0111483} {arXiv:astro-ph/0111483}
  \BibitemShut {NoStop}%
\bibitem [{\citenamefont {Gogoberidze}\ \emph {et~al.}(2007)\citenamefont
  {Gogoberidze}, \citenamefont {Kahniashvili},\ and\ \citenamefont
  {Kosowsky}}]{Gogoberidze:2007an}%
  \BibitemOpen
  \bibfield  {author} {\bibinfo {author} {\bibfnamefont {G.}~\bibnamefont
  {Gogoberidze}}, \bibinfo {author} {\bibfnamefont {T.}~\bibnamefont
  {Kahniashvili}},\ and\ \bibinfo {author} {\bibfnamefont {A.}~\bibnamefont
  {Kosowsky}},\ }\bibfield  {title} {\bibinfo {title} {{The Spectrum of
  Gravitational Radiation from Primordial Turbulence}},\ }\href
  {https://doi.org/10.1103/PhysRevD.76.083002} {\bibfield  {journal} {\bibinfo
  {journal} {Phys. Rev. D}\ }\textbf {\bibinfo {volume} {76}},\ \bibinfo
  {pages} {083002} (\bibinfo {year} {2007})},\ \Eprint
  {https://arxiv.org/abs/0705.1733} {arXiv:0705.1733 [astro-ph]} \BibitemShut
  {NoStop}%
\bibitem [{\citenamefont {Caprini}\ \emph {et~al.}(2008)\citenamefont
  {Caprini}, \citenamefont {Durrer},\ and\ \citenamefont
  {Servant}}]{Caprini:2007xq}%
  \BibitemOpen
  \bibfield  {author} {\bibinfo {author} {\bibfnamefont {C.}~\bibnamefont
  {Caprini}}, \bibinfo {author} {\bibfnamefont {R.}~\bibnamefont {Durrer}},\
  and\ \bibinfo {author} {\bibfnamefont {G.}~\bibnamefont {Servant}},\
  }\bibfield  {title} {\bibinfo {title} {{Gravitational wave generation from
  bubble collisions in first-order phase transitions: An analytic approach}},\
  }\href {https://doi.org/10.1103/PhysRevD.77.124015} {\bibfield  {journal}
  {\bibinfo  {journal} {Phys. Rev. D}\ }\textbf {\bibinfo {volume} {77}},\
  \bibinfo {pages} {124015} (\bibinfo {year} {2008})},\ \Eprint
  {https://arxiv.org/abs/0711.2593} {arXiv:0711.2593 [astro-ph]} \BibitemShut
  {NoStop}%
\bibitem [{\citenamefont {Caprini}\ \emph
  {et~al.}(2009{\natexlab{a}})\citenamefont {Caprini}, \citenamefont {Durrer},\
  and\ \citenamefont {Servant}}]{Caprini:2009yp}%
  \BibitemOpen
  \bibfield  {author} {\bibinfo {author} {\bibfnamefont {C.}~\bibnamefont
  {Caprini}}, \bibinfo {author} {\bibfnamefont {R.}~\bibnamefont {Durrer}},\
  and\ \bibinfo {author} {\bibfnamefont {G.}~\bibnamefont {Servant}},\
  }\bibfield  {title} {\bibinfo {title} {{The stochastic gravitational wave
  background from turbulence and magnetic fields generated by a first-order
  phase transition}},\ }\href {https://doi.org/10.1088/1475-7516/2009/12/024}
  {\bibfield  {journal} {\bibinfo  {journal} {JCAP}\ }\textbf {\bibinfo
  {volume} {12}},\ \bibinfo {pages} {024}},\ \Eprint
  {https://arxiv.org/abs/0909.0622} {arXiv:0909.0622 [astro-ph.CO]}
  \BibitemShut {NoStop}%
\bibitem [{\citenamefont {Caprini}\ \emph
  {et~al.}(2009{\natexlab{b}})\citenamefont {Caprini}, \citenamefont {Durrer},
  \citenamefont {Konstandin},\ and\ \citenamefont {Servant}}]{Caprini:2009fx}%
  \BibitemOpen
  \bibfield  {author} {\bibinfo {author} {\bibfnamefont {C.}~\bibnamefont
  {Caprini}}, \bibinfo {author} {\bibfnamefont {R.}~\bibnamefont {Durrer}},
  \bibinfo {author} {\bibfnamefont {T.}~\bibnamefont {Konstandin}},\ and\
  \bibinfo {author} {\bibfnamefont {G.}~\bibnamefont {Servant}},\ }\bibfield
  {title} {\bibinfo {title} {{General Properties of the Gravitational Wave
  Spectrum from Phase Transitions}},\ }\href
  {https://doi.org/10.1103/PhysRevD.79.083519} {\bibfield  {journal} {\bibinfo
  {journal} {Phys. Rev. D}\ }\textbf {\bibinfo {volume} {79}},\ \bibinfo
  {pages} {083519} (\bibinfo {year} {2009}{\natexlab{b}})},\ \Eprint
  {https://arxiv.org/abs/0901.1661} {arXiv:0901.1661 [astro-ph.CO]}
  \BibitemShut {NoStop}%
\bibitem [{\citenamefont {Behnel}\ \emph {et~al.}(2011)\citenamefont {Behnel},
  \citenamefont {Bradshaw}, \citenamefont {Citro}, \citenamefont {Dalcin},
  \citenamefont {Seljebotn},\ and\ \citenamefont {Smith}}]{cython}%
  \BibitemOpen
  \bibfield  {author} {\bibinfo {author} {\bibfnamefont {S.}~\bibnamefont
  {Behnel}}, \bibinfo {author} {\bibfnamefont {R.}~\bibnamefont {Bradshaw}},
  \bibinfo {author} {\bibfnamefont {C.}~\bibnamefont {Citro}}, \bibinfo
  {author} {\bibfnamefont {L.}~\bibnamefont {Dalcin}}, \bibinfo {author}
  {\bibfnamefont {D.~S.}\ \bibnamefont {Seljebotn}},\ and\ \bibinfo {author}
  {\bibfnamefont {K.}~\bibnamefont {Smith}},\ }\bibfield  {title} {\bibinfo
  {title} {Cython: The best of both worlds},\ }\href
  {https://doi.org/10.1109/MCSE.2010.118} {\bibfield  {journal} {\bibinfo
  {journal} {Computing in Science Engineering}\ }\textbf {\bibinfo {volume}
  {13}},\ \bibinfo {pages} {31} (\bibinfo {year} {2011})}\BibitemShut {NoStop}%
\bibitem [{\citenamefont {Dalc{\'\i}n}\ \emph {et~al.}(2005)\citenamefont
  {Dalc{\'\i}n}, \citenamefont {Paz},\ and\ \citenamefont {Storti}}]{mpi4py}%
  \BibitemOpen
  \bibfield  {author} {\bibinfo {author} {\bibfnamefont {L.}~\bibnamefont
  {Dalc{\'\i}n}}, \bibinfo {author} {\bibfnamefont {R.}~\bibnamefont {Paz}},\
  and\ \bibinfo {author} {\bibfnamefont {M.}~\bibnamefont {Storti}},\
  }\bibfield  {title} {\bibinfo {title} {Mpi for python},\ }\href
  {https://doi.org/https://doi.org/10.1016/j.jpdc.2005.03.010} {\bibfield
  {journal} {\bibinfo  {journal} {Journal of Parallel and Distributed
  Computing}\ }\textbf {\bibinfo {volume} {65}},\ \bibinfo {pages} {1108}
  (\bibinfo {year} {2005})}\BibitemShut {NoStop}%
\bibitem [{\citenamefont {Harris}\ \emph {et~al.}(2020)\citenamefont {Harris}
  \emph {et~al.}}]{2020NumPy-Array}%
  \BibitemOpen
  \bibfield  {author} {\bibinfo {author} {\bibfnamefont {C.~R.}\ \bibnamefont
  {Harris}} \emph {et~al.},\ }\bibfield  {title} {\bibinfo {title} {Array
  programming with {NumPy}},\ }\href
  {https://doi.org/10.1038/s41586-020-2649-2} {\bibfield  {journal} {\bibinfo
  {journal} {Nature}\ }\textbf {\bibinfo {volume} {585}},\ \bibinfo {pages}
  {357} (\bibinfo {year} {2020})}\BibitemShut {NoStop}%
\bibitem [{\citenamefont {Virtanen}\ \emph {et~al.}(2020)\citenamefont
  {Virtanen} \emph {et~al.}}]{2020SciPy-NMeth}%
  \BibitemOpen
  \bibfield  {author} {\bibinfo {author} {\bibfnamefont {P.}~\bibnamefont
  {Virtanen}} \emph {et~al.},\ }\bibfield  {title} {\bibinfo {title} {{{SciPy}
  1.0: Fundamental Algorithms for Scientific Computing in Python}},\ }\href
  {https://doi.org/10.1038/s41592-019-0686-2} {\bibfield  {journal} {\bibinfo
  {journal} {Nature Methods}\ }\textbf {\bibinfo {volume} {17}},\ \bibinfo
  {pages} {261} (\bibinfo {year} {2020})}\BibitemShut {NoStop}%
\bibitem [{\citenamefont {Dahl}(2021{\natexlab{a}})}]{dahl_jani_2021_5786090}%
  \BibitemOpen
  \bibfield  {author} {\bibinfo {author} {\bibfnamefont {J.}~\bibnamefont
  {Dahl}},\ }\href {https://doi.org/10.5281/zenodo.5786090} {\bibinfo {title}
  {Simulation code for fluids in two dimensions}} (\bibinfo {year}
  {2021}{\natexlab{a}})\BibitemShut {NoStop}%
\bibitem [{\citenamefont {Dahl}(2021{\natexlab{b}})}]{dahl_jani_2021_5786442}%
  \BibitemOpen
  \bibfield  {author} {\bibinfo {author} {\bibfnamefont {J.}~\bibnamefont
  {Dahl}},\ }\href {https://doi.org/10.5281/zenodo.5786442} {\bibinfo {title}
  {{Simulation code for fluids in two dimensions - density and velocity
  movies}}} (\bibinfo {year} {2021}{\natexlab{b}})\BibitemShut {NoStop}%
\bibitem [{\citenamefont {Dahl}(2021{\natexlab{c}})}]{dahl_jani_2021_5786393}%
  \BibitemOpen
  \bibfield  {author} {\bibinfo {author} {\bibfnamefont {J.}~\bibnamefont
  {Dahl}},\ }\href {https://doi.org/10.5281/zenodo.5786393} {\bibinfo {title}
  {{Simulation code for fluids in two dimensions - vorticity movie}}} (\bibinfo
  {year} {2021}{\natexlab{c}})\BibitemShut {NoStop}%
\end{thebibliography}%

\end{document}